\documentclass[10pt]{article}
\usepackage[utf8]{inputenc}
\date{24/05/19}
\usepackage{amsmath}
\usepackage{amsfonts}
\usepackage{amssymb}
\usepackage{extsizes}
\usepackage{graphicx}
\usepackage[left=2cm,right=2cm,top=2cm,bottom=4cm]{geometry}

\date{\today}
\begin{document}
\title{{\bf{Spectral  Distance on Lorentzian Moyal Plane}}}
\author{
{\bf {\normalsize Anwesha Chakraborty}$^{a}$
\thanks{anwesha@bose.res.in}},
{\bf {\normalsize Biswajit Chakraborty}$^{b}$
\thanks{biswajit@bose.res.in}}
\\
$^{a,b}$ {\normalsize Department of Theoretical Sciences}\\
{\normalsize S.N. Bose National Centre for Basic Sciences}\\
{\normalsize JD Block, Sector III, Salt Lake, Kolkata 700106, India}\\}
\maketitle
\begin{abstract}
We present here a completely operatorial approach, using Hilbert-Schmidt operators, to compute spectral distances between time-like separated \textquotedblleft events \textquotedblright , associated with the pure states of the algebra describing the Lorentzian Moyal plane, using the axiomatic framework given by \cite{Franco1, Franco2}. The result shows no deformations of non-commutative origin, as in the Euclidean case.
\end{abstract}

\section{Introduction}
%Non-commutative geometry is the mathematical framework where we work with the algebra of functions defined on the   space instead of working with the space itself. Though we can trace back the geometrical space from the algebra for commutative manifold aka Gelfand, Naimark , but this is not true for non-commutative (NC) spaces. In general we should say one can  work with the algebra but there is no information of space whatsoever. It was first observed by Alain Connes' et.al in their functional analytic approach to Non-commutative geometry that the metric information of a compact Riemannian  manifold is encoded in the set of three components called Spectral Triple $(\mathcal{A},\mathcal{H},\mathcal{D})$ where $\mathcal{A}$ is a $*$ algebra of bounded operators on Hilbert space $\mathcal{H}$ and $\mathcal{D}$ is the unbounded Dirac operator defined on it. The space of  elements belonging to the algebra is isomorphic to the set of points in the corresponding manifold. \\ \\
%The formulation of  Lorentzian spectral triple is needed as because causality of time-like events are not reflected in Euclidean Non-commutative geometry.  
%It becomes evident that the possible Dirac operator no more remains self-adjoint on the Hilbert space. The underlined space is now a Krein space where the defined Dirac operator is a Krein self-adjoint. 
In his formulation of Non-commutative geometry (NCG) \cite{Connes} and its subsequent application to standard model of particle physics \cite{Suij,Lizzi}, Alain Connes has essentially dealt with spaces with Euclidean signature i.e. Riemannian manifold. This feature of his formulation has remained a sort of a bottle-neck in  the further development in its application  and eventual reconciliation with the realistic nature of our space-time, which as we all know to be  as a manifold with Lorentzian signature. Attempts are being made for quite some time now  \cite{Eli,Wd,Kv} and  people have devised various ways to circumvent  this problem like the so called \textquotedblleft Wick rotation\textquotedblright \cite{Lizzi} etc. However, the unanimous opinion among the practitioners of NCG and its application to standard model is that one needs to confront this issue of Lorentzian signature head-on. Recent activities  in this direction indicates that there is still no consensus in the literature about its axiomatic formulation \cite{Kopf,Barrett,Sitarz,Stro,Parf,Moretti,Franco1}. Despite the fact, we would like to follow the work of \cite{Franco1,Franco2}, in our preliminary attempt to compute the spectral distance between a pair of time-like separated events associated with pure states in Lorentzian Moyal plane  and show that this axiomatic formulations of \cite{Franco1,Franco2}, serves our purpose quite adequately. This work can be thought of as a sequel of our earlier work on Euclidean Moyal plane \cite{Chaoba} and  here too we are employing Hilbert-Schmidt  operator formulation of NC quantum mechanics \cite{Scholtz}. \\ \\
The paper is organised as follows:   In section 2  after a brief recapitulation of Hilbert-Schmidt approach in quantum mechanics we make use of Fock-Bergman coherent state basis to represent the states of the Hilbert space. The  functional differentiation approach is then  employed for the extremisation of various functionals arising here naturally  in the derivation of spectral distance between a pair of pure states on Euclidean Moyal plane. Section 3 is the brief outline of the construction of Lorentzian spectral triple and emergence of algebraic ball condition as a result of causality following  \cite{Franco1,Franco2}. The next section consists of the explicit derivation of Dirac operator and Ball condition for two dimensional Lorentzian commutative and non-commutative manifold with (-,+) signature, followed by the calculation of distances in functional derivative approach for the respective manifolds, where we have also discussed the Poincare invariance of distance and non-invariance of \textquotedblleft vacuum\textquotedblright under Lorentz boost. Finally we conclude in section 5.      
\section{Recapitulation  of the Computation of Spectral Distance on Moyal Plane with Euclidean Signature}
Moyal plane is defined through a commutator relation \begin{equation}
[\hat{x}_1,\hat{x}_2]=i\theta;\,\,\,\theta>0\label{algebra}
\end{equation}
 satisfied by the operator valued coordinates $\hat{x}_1$ and $\hat{x}_2$. As it stands, this commutator algebra does not specify the signature of the underlying commutative space (i.e. the one obtained in the commutative limit), in the sense that  it can correspond to either Euclidean or Lorentzian signature, as can be seen easily from the fact that this noncommutative algebra  (\ref{algebra}) is invariant under both SO(2) and SO(1,1) transformation. However, since Connes' formulation of NCG \cite{Connes2}  requires in the commutative case, the spin manifold to be Euclidean, the initial works on the computation of spectral distances between pure states in the Moyal plane \cite{Mart} and fuzzy sphere \cite{Liz}  were restricted to the Euclidean signatures only, in the sense that the respective commutative limits yield the corresponding manifold with the associated metrics. In particular , in \cite{Chaoba} computations were carried out in a complete operatorial level, using Hilbert-Schmidt operator, which requires  no use of any star product,  thereby bypassing any ambiguities that may result from there \cite{Prasad}. In the present paper, as mentioned above, we would like to extend our computation of similar distances for the Lorentzian signature as well. For that it will be advantageous to begin by recapitulating briefly the computation of distances with Euclidean signature as in \cite{Chaoba}. but, in contrast to \cite{Chaoba}, we shall employ the Fock-Bergman coherent states  to represent state vectors belonging to the Hilbert space and eventually to extremise the ball functional. It will make the computations much more transparent and can be appreciated by any potential reader.  This will help us to set up the basic frame work and introduce the necessary notations, so that the same can be emulated and adapted in the Lorentzian case. In fact as we shall see that there are several points of contacts between these two cases with respective similarities and dissimilarities.\\ \\ \\
The spectral distance on Moyal plane is computed by employing the spectral triple $(\mathcal{A},\mathcal{H},\mathcal{D})$  where
\begin{equation}
{A}=\mathcal{H}_q, \,\,\, \mathcal{H}=\mathcal{H}_c\otimes \mathbb{C}^2,\,\,\,\mathcal{D}=\sqrt{\frac{2}{\theta}}\begin{pmatrix}
0&\hat{b}^{\dagger}\\\hat{b}&0 
\end{pmatrix}\label{o}
\end{equation}
and $\mathcal{H}_q$ is the space of Hilbert-Schmidt (HS) operators acting on   \textquotedblleft configuration space  \textquotedblright  $\,\mathcal{H}_c $ - which is also a Hilbert space, and defined as,
\begin{equation}
\mathcal{H}_c=\textrm{Span}\left\{|n\rangle =\frac{(\hat{b}^{\dagger})^n}{\sqrt{n!}}|0\rangle\right\}; \,\,\,\, \hat{b}=\frac{\hat{x}_1+i\hat{x}_2}{\sqrt{2\theta}}\,\,\,\, \textrm{and}\,\, \hat{b}|0\rangle = 0. \label{k}
\end{equation}
This furnishes a representation of just the above coordinate algebra (\ref{algebra}) and is basically isomorphic to the Hilbert space of 1D  harmonic oscillator. $\mathcal{H}_q$ is  equipped with the inner product $(\psi,\phi)=tr_{\mathcal{H}_c}(\psi^{\dagger}\phi)$ and as a part of the definition the associated HS norm defined as $\Vert\psi\Vert_{HS}:= \sqrt{tr_{\mathcal{H}_c}(\psi^{\dagger}\psi)} <\infty$ i.e. fulfills finiteness property. In other words, a generic HS operator (denoted sometimes by a round ket $|.)$, in contrast to angular $|.\rangle$ notations used  for the elements of $\mathcal{H}_c$ (\ref{k})) $\psi :=|\psi)\in\mathcal{H}_q$ can be expanded in the Fock basis (\ref{k}) as,
\begin{equation}
|\psi) = \sum_{m,n}C_{m,n}|m\rangle\langle n| \label{ju}
\end{equation}
as $\mathcal{H}_q$ can be identified as $\mathcal{H}_c\otimes\tilde{\mathcal{H}_c}$ with $\tilde{\mathcal{H}_c}$ being the dual Hilbert space. Now making use of the identity $|0\rangle\langle 0| = :e^{-\hat{b}^{\dagger}\hat{b}}:$, where the colons represent normal ordering of the operators \cite{Itz},  one can see that $\psi$ can be recast as a polynomial algebra generated by $\hat{b},\hat{b}^{\dagger}$ as,
\begin{equation}
|\psi) =\sum_{m,n} C_{mn}\frac{\hat{b}^{\dagger m}}{\sqrt{m!}}|0\rangle\langle 0| \frac{\hat{b}^n}{\sqrt{n!}} = \sum_{m,n,l}C_{mn}\frac{1}{l!}\frac{(-1)^l}{\sqrt{m!n!}} (\hat{b}^{\dagger} )^{m+l}(\hat{b})^{n+l} \,\,\,\,\in \mathcal{H}_q , \label{poly}
\end{equation}
which is automatically in the normal-ordered form. Note that  $\Vert\psi\Vert_{HS}^2 =\sum_{m,n}|C_{mn}|^2 < \infty $, as required by very definition of  HS operator. This  $\mathcal{H}_q$, whose generic forms are like in (\ref{ju}), has the structure of an algebra. This allows us to identify the algebra $\mathcal{A}$ to $\mathcal{H}_q$ itself: $\mathcal{A}=\mathcal{H}_q$ in (\ref{o}). It acts on $\mathcal{H}_c\otimes\mathbb{C}^2$ through the diagonal representation $\pi(a):= \textrm{diag}(a,a)$ from left so that this Hilbert space $\mathcal{H}_c\otimes\mathbb{C}^2$ can  be regarded as the left module of the algebra. Also note that the algebra $\mathcal{A}=\mathcal{H}_q$ is a dense subspace of $\mathcal{B}(\mathcal{H}_c)$, where this can be identified with a $C^*$ - algebra with $*$- operation denoting hermitian conjugation which play the role of involution. Finally  $\mathcal{D}$  in (\ref{o}) is the Dirac operator  whose construction has been reviewed in \cite{Chaoba}. The point that we would like to emphasise is that the structure of the Dirac operator turned out to be  such that  it can  also act on the Hilbert space  $\mathcal{H}_c\otimes\mathbb{C}^2$ from left.   Consequently, the Hilbert space in the spectral triple is taken to be $\mathcal{H}_c\otimes\mathbb{C}^2$, rather than $\mathcal{H}_q\otimes\mathbb{C}^2$- as would have been expected from the corresponding commutative case where  it is taken to be of the form $L^2(\mathbb{R}^2)\otimes\mathbb{C}^2$.  \\ \\
Here one has two kinds of choices of  states viz.  the normalised coherent state 
\begin{equation}
|z\rangle =e^{-\bar{z}\hat{b}+z\hat{b}^{\dagger}}|0\rangle \in\mathcal{H}_c ;\,\,\, \langle z|z\rangle =1 \label{coh}
\end{equation}
 and the   states $|n\rangle$, as  in  (\ref{k}) . Correspondingly,  one can introduce pure states $\rho_z:=|z\rangle\langle z|,\,\,$ and the so called \textquotedblleft harmonic oscillator \textquotedblright  states $ \rho_n:=|n\rangle\langle n| \in \mathcal{H}_q$ which are linear functionals of unit  norm,  acting on the algebra  $\mathcal{A}=\mathcal{H}_q$ as \begin{equation}
\rho_z(\hat{a})=\textrm{tr}_{\mathcal{H}_c}(\rho_z\hat{a})=(\rho_z,\hat{a})= \langle z|\hat{a}|z\rangle\,\, ; \rho_n(\hat{a})=tr_{\mathcal{H}_c}(\rho_n  \hat{a}) =\langle n|\hat{a}|n\rangle  \label{u}
\end{equation}
These pure states $\rho_z $ and $\rho_n$ can also be regarded as density matrices as viewed from  $\mathcal{H}_c$. In the spirit of Gelfand and Naimark, here too one can associate $\rho_z$, in particular, with the point having the complex coordinate $z$ in the Argand diagram with the latter being viewed as smeared Moyal plane. Spectral  distance between a generic pair of states $\rho_z$  and $\rho_{w}$ (in this paper we shall be considering only pure states built out of coherent states) \textit{a la} Connes'  is given  by 
\begin{equation}
d(\rho_z,\rho_w)=\sup_{\hat{a}\in \mathcal{B}}\{|\rho_z(\hat{a})-\rho_w(\hat{a})| \}\label{hg}
\end{equation}
 where $\mathcal{B}$ is the Ball defined as 
  \begin{equation}
 \mathcal{B}=\{\hat{a}:\Vert[\mathcal{D},\pi(\hat{a})]\Vert_{op}\leq 1 \}\label{ki}
 \end{equation}
 Given the structure of $\mathcal{D}$ in  (\ref{o}) the Ball condition can be written equivalently and more compactly as 
\begin{equation}
\mathcal{B}=\left\{\hat{a}\,:\, \Vert[\hat{b},\hat{a}]\Vert_{op}=\Vert[\hat{b}^{\dagger},\hat{a}]\Vert_{op}\leq \sqrt{\frac{\theta}{2}}\right\}\label{ballcondition}
\end{equation}
As has been noted in  \cite{Chaoba} that the optimal algebra element $a_s\in \mathcal{A}$ for which the supremum is attained  in (\ref{hg}) , yielding the distance i.e. 
\begin{equation}
d(\rho_z,\rho_w)=|\rho_z(\hat{a}_s)-\rho_w(\hat{a}_s)| \label{hp}
\end{equation}
must also saturate the Ball condition (\ref{ballcondition}) i.e. it should satisfy 
\begin{equation}
\Vert [\hat{b},\hat{a}_s]\Vert_{op}=\Vert [\hat{b}^{\dagger},\hat{a}_s]\Vert_{op}=\sqrt{\frac{\theta}{2}}\label{A3}
\end{equation}
Further following  \cite{Mart2}, we know that the search of such an optimal algebra element can be restricted to hermitian algebra elements only i.e. we require $\hat{a}_s^{\dagger}= \hat{a}_s$.
To that end we take up the simplified Ball condition (\ref{ballcondition}) and first try to compute $\Vert[\hat{b},\hat{a}]\Vert_{op}^2$. Making use of the Fock-Bergman basis (\ref{A67}) and the associated resolution of identity (\ref{A68}) (see Appendix-A), one gets,
\begin{equation}
||[\hat{b},\hat{a}]||_{op}^2 = \sup_{||\psi||=1}\langle\psi|[\hat{b},\hat{a}]^{\dagger}[\hat{b},\hat{a}]|\psi\rangle =\sup_{||\psi||=1} \int d\mu({z,\bar{z}})d\mu(w,\bar{w})\psi^*(z)\psi(\bar{w})\langle \bar{z}|[\hat{b},\hat{a}]^{\dagger}[\hat{b},\hat{a}] |w\rangle \label{lom}
\end{equation}
A simple way to compute the supremum is to employ the  method of Lagrange's undetermined multiplier for extremizing the R.H.S of the above equation subject to the constraint $\langle\psi|\psi\rangle =1$. We therefore extremise  the functional
\begin{equation*}
\textbf{B}[\psi^*(z),\psi(\bar{z});\lambda] := \int d\mu({z,\bar{z}})d\mu(w,\bar{w})\psi^*(z)\psi(\bar{w})\langle \bar{z}|[\hat{b},\hat{a}]^{\dagger}[\hat{b},\hat{a}] |w\rangle-\lambda\int d\mu(\bar{z},z) [\psi^*(z)\psi(\bar{z})-1]
\end{equation*}
with $\lambda$ being a real valued Lagrange's multiplier enforcing the constraint $\langle\psi|\psi\rangle =1$. By varying $\textbf{B}[\psi^*(z),\psi(\bar{z});\lambda]$  with respect to $\psi(\bar{z}),\psi^*(z)$ we get the following pair of equivalent equations,
\begin{align}
\langle\psi|[\hat{b},\hat{a}]^{\dagger}[\hat{b},\hat{a}]|z\rangle &=\lambda \langle\psi|z\rangle\nonumber\\
\langle\bar{z}|[\hat{b},\hat{a}]^{\dagger}[\hat{b},\hat{a}]|\psi\rangle &=\lambda \langle\bar{z}|\psi\rangle.\label{A9}
\end{align}
In carrying out this functional differentiation we have used the basic relations like  (\ref{A70}) and (\ref{A71}) (see  Appendix-A). Now since (\ref{A9}) holds for arbitrary  $|\psi\rangle$ and $|z\rangle$, we  have the matrix elements of the operator $[\hat{b},\hat{a}]^{\dagger}[\hat{b},\hat{a}]$  between  any pair of states  $|\psi\rangle$ and $|\phi\rangle$  must satisfy $\langle\psi|[\hat{b},\hat{a}]^{\dagger}[\hat{b},\hat{a}]|\phi\rangle =\lambda\langle\psi|\phi\rangle;\,\,\, \forall\,\,\, |\psi\rangle,|\phi\rangle\in\mathcal{H}_c$. Consequently, we must have the following operator identity holding: 
\begin{equation}
[\hat{b},\hat{a}]^{\dagger}[\hat{b},\hat{a}]=\lambda\label{bp}
\end{equation} 
The structure of this equation apparently indicates that $\hat{a}$ can only be related linearly to $\hat{b}$ and $\hat{b}^{\dagger}$, so that a real number is produced in the R.H.S as can be seen from the general structure of the algebra element $\hat{a}$ (\ref{poly}). This, however, is not entirely true! Indeed, as has been shown in \cite{Chaoba} that there exists finite dimensional matrix solution for $\hat{a}$,  so that the positive operator $[\hat{b},\hat{a}]^{\dagger}[\hat{b},\hat{a}]$ is only a diagonal matrix, which is not proportional to the unit matrix, so that one can identify states  $|\psi_i\rangle \in \mathcal{H}_c$ which corresponds to local extrema and with the associated eigen value $\lambda_i\geq 0$ as $[\hat{b},\hat{a}]^{\dagger}[\hat{b},\hat{a}]|\psi_i\rangle =\lambda_i|\psi_i\rangle$ . Eventually, one can just read-off the maximum eigenvalue $\lambda_{max}$ and identify $\Vert[\hat{b},\hat{a}]\Vert_{op}=\sqrt{\lambda_{max}}$ (see Appendix B). These finite dimensional matrix solutions of $\hat{a}$ which should result in a distance with $\lambda_{max}=\frac{\theta}{2}$ are employed to compute distances between the \textquotedblleft harmonic oscillator \textquotedblright states $\rho_n$ and $\rho_m$: $d(\rho_n,\rho_m)$, where they serve as an optimal element. Here, on the contrary, we are interested in computing  the distance between a pair of  coherent states $\rho_z$  and $\rho_w$ (\ref{hg}), where we can see from the very definition (\ref{coh}) of coherent states itself that we need a non-trivial infinite dimentional solution for '$\hat{a}$ ' which should result in  a distance, that have a  manifest invariance property under ISO(2) \cite{Chaoba}. Further, in this case, we  can also provide an upper bound to the distance (\ref{hg}). It will be useful to  recapitulate this derivation  here very briefly. Here one introduces a one- parameter family of pure states $\rho_{\mu z}:=|\mu z\rangle\langle\mu z |$ with $\mu\in[0,1]$ being a real parameter, interpolating between $\rho_0$ and $\rho_z$ (Note that, we have taken, without loss of generality, $w=0$, by invoking the above mentioned ISO(2) invariance). We then re-write the expression occurring in the RHS of (\ref{hg})   as,
\begin{equation}
|\rho_z(\hat{a})-\rho_0(\hat{a})|=\left|\int_0^1 d\mu \frac{d\rho_{\mu z}(\hat{a})}{d\mu}\right| \leq \int_0^1 d\mu \left|\frac{d\rho_{\mu z}(\hat{a})}{d\mu}\right| = \int_0^1 d\mu \left|\bar{z}\rho_{\mu z}[\hat{b},\hat{a}]+z\rho_{\mu z}[\hat{b},\hat{a}]^{\dagger}\right|\label{B1}
\end{equation}
By applying Cauchy-Schwartz inequality, this inequality can further be simplified as ,
\begin{equation}
|\rho_z(\hat{a})-\rho_0(\hat{a})| \leq \sqrt{2}|z|\int_0^1 d\mu \sqrt{\left|\rho_{\mu z}[\hat{b},\hat{a}]\right|^2+\left|\rho_{\mu z}[\hat{b},\hat{a}]^{\dagger}\right|^2} \leq 2|z|\sqrt{ \Vert [\hat{b},\hat{a}]^{\dagger}[\hat{b},\hat{a}]\Vert_{op}} = 2|z|\,\,\Vert [\hat{b},\hat{a}\Vert_{op}\label{B2}
\end{equation} 
Finally  making use of  the ball condition (\ref{A3}), we get  the desired upper bound on the distance as,
\begin{equation}
d(\rho_z,\rho_0) \leq \sqrt{2\theta} |z|\label{B3}
\end{equation}
Equivalently by re-introducing  z and $w$ variables again by invoking ISO(2) symmetry we have,
\begin{equation}
d(\rho_z,\rho_w) \leq \sqrt{2\theta} |z-w|\label{B4}
\end{equation}  
We can now take  the ansatz for $\hat{a}$ as $\hat{a}=\xi\hat{b}+\bar{\xi}\hat{b}^{\dagger}$ with $\xi\in\mathbb{C}$ (as the algebra element should be hermitian). We  have from (\ref{bp}),  $|\xi|^2 =\lambda$. To  evaluate the value of $\lambda$ we make use of  (\ref{ballcondition}), (\ref{lom})and (\ref{bp})  to have,
\begin{equation*}
\Vert[\hat{b},\hat{a}]\Vert_{op}= \sup_{\Vert\psi\Vert=1}\sqrt{\langle\psi|[\hat{b},\hat{a}]^{\dagger}[\hat{b},\hat{a}]|\psi\rangle} =\sqrt{\lambda} = \sqrt{\frac{\theta}{2}}
\end{equation*}
So the value of $\lambda$ for which the Ball reaches its supremum value is $\lambda=\frac{\theta}{2}$. So the optimal algebra element $\hat{a}_s$ lies within the one parameter family given by the phase $\alpha$ as,
\begin{equation}
\hat{a}_s \in  \left\{\sqrt{\frac{\theta}{2}}(\hat{b}e^{-i\alpha}+\hat{b}^{\dagger}e^{i\alpha});\,\, 0\leq\alpha<2\pi\right\}\label{set}
\end{equation}
where we have taken  $\xi =e^{-i\alpha} \sqrt{\frac{\theta}{2}}$.
One can further corroborate this by observing that the matrix element of $[\hat{b},\hat{a}_s]$ with  $\hat{a}_s$ given by (\ref{set}),  in the Bergman-Fock basis (\ref{A67}) is obtained as, 
\begin{equation}
_B\langle \bar{z}^{\prime}|[\hat{b},\hat{a}]|z\rangle_B=\sqrt{\frac{\theta}{2}}e^{i\alpha}e^{\bar{z}^{\prime}z}
 \end{equation}
 where we recognize the occurrence of  Dirac's delta function in this Bergman-Fock space: $e^{\bar{z}^{\prime}z}=\delta(\bar{z}^{\prime},z) $  (see (\ref{A71}  in Appendix A) ,  so that 
$[\hat{b},\hat{a}_s]$  is proportional to  the unit matrix, when written in this continuous Bergman-Fock basis, enabling one to just read-off the operator norm. It should be noted here that, in contrast to the finite dimensional matrix $\hat{a}_s$ (see (\ref{form3}) in Appendix-B ) , the $\hat{a}_s$  in (\ref{set}) yields $\lambda=\frac{\theta}{2}$ as the unique eigenvalue of $[\hat{b},\hat{a}]^{\dagger}[\hat{b},\hat{a}]$, which is now infinitely degenerate in the eigen-basis provided by whole continuum  set of Fock-bergman coherent states $|z\rangle$ (\ref{A67}) and therefore is independent of $z$. This plays a  vital role in making the distances invariant under translation, in contrast to the distance between the harmonic oscillator states . For example $d(\rho_{n+1},\rho_n)\neq d(\rho_{n+2},\rho_{n+1})$ \cite{Chaoba,Mart3}.  One can also see that with this identification of $\lambda$, the upper bound on the distance indeed takes the form of (\ref{B4}).
 Now substituting (\ref{set}) in   (\ref{hg}) the distance is essentially found to be  given by    
 \begin{equation}
d(\rho_z,\rho_w)=\sup_{\hat{a}\in\mathcal{A}}\,\,\left|\langle z|\hat{a}_s|z\rangle -\langle w|\hat{a}_s|w\rangle\right|= \sqrt{\frac{\theta}{2}}\,\, \sup_{\alpha} \,\,|(z-w)e^{-i\alpha} +(\bar{z}-\bar{w})e^{i\alpha}|
\end{equation}
Note that as $\rho_z,\,\rho_w$ are normalised  pure state density matrices, we have to use the normalised coherent state basis here for distance calculation as defined in (\ref{coh}), instead of Fock-Bergman basis\footnote{To avoid cluttering of indices, here after we shall use the same notation $|z\rangle$ to indicate both the normalised coherent state (\ref{coh}) and the Fock-Bergman(un-normalised basis) $|z\rangle_B$ in (\ref{A67}). It should be clear from the context, which is being used and should not cause any   confusion. Particularly, $|z\rangle$ occuring in density matrix $\rho_z=|z\rangle\langle z| $ is clearly the normalised one, otherwise, it is the latter one.}.
We can now parametrize the complex number by polar decomposition as $z-w=|z-w|e^{i\beta},\,\,\bar{z}-\bar{w}=|z-w|e^{-i\beta}$. The optimal algebra element for which the supremum value is attained  can easily be recognised to be the one for which $\alpha=\beta$  yielding the desired  distance between two arbitrary pure states,
\begin{equation}
d(\rho_z,\rho_w) = \sqrt{2\theta}\,\,|z-w|\label{mp}
\end{equation}
which precisely is the upper bound (\ref{B4}). This inequality is therefore saturated by $\hat{a}_s$ (\ref{set}). Splitting dimensionless complex coordinates $z$ and $w$ into real and imaginary parts involving dimension-ful coordinates, in analogy with the splitting of $\hat{b}$ in (\ref{k}) as 
\begin{equation}
z=\frac{1}{\sqrt{2\theta}}(x_1+ix_2)\,\,\,\textrm{and}\,\,\, w=\frac{1}{\sqrt{2\theta}} (y_1+iy_2)\label{basis}
\end{equation} 
we can rewrite (\ref{mp}) as,
\begin{equation}
d(\rho_z,\rho_w)=\sqrt{(x_1-y_1)^2+(x_2-y_2)^2}\label{mmp}
\end{equation}
reproducing the usual Euclidean distance having the complete ISO(2) symmetry  as in\cite{Mart,Chaoba}, with no NC corrections. Before we conclude this section, we would like to point out that the form of the algebra element, given in (\ref{set}), are not the elements of the algebra:  $a_s \notin \mathcal{A}$. But these elements can be regarded to belong to an enlarged algebra, called multiplier algebra (For details see \cite{Mart,Bimo}). We shall not elaborate on this any further here and just mention that  at least in our case the operator product of $\rho_z$ and $\hat{a}_s$ in (\ref{set}) is indeed a HS operator, as follows from the fact that $\Vert \rho_z\hat{a}_s\Vert_{HS}^2= tr_{\mathcal{H}_c}((\rho_z\hat{a}_s)^{\dagger}(\rho_z\hat{a}_s))=tr_{\mathcal{H}_c}(\hat{a}_s\rho_z\hat{a}_s) < \infty$. Furthermore, this will be a recurrent feature in the rest of the paper i.e. we will have to consider the multiplier algebra in the Lorentzian case also, which we shall deal with in the sequel.
\section{Lorentzian Spectral Triple}
In this section we would like to use a formulation required in the computation of distances between a pair of time-like separated events in commutative and non-commutative  in Lorentzian  plane. Although, to the best of our knowledge, a  general consensus  regarding its recently proposed various axiomatic frameworks is still lacking, we nevertheless try here to follow  \cite{Franco1,Franco2,Franco3}  and adapt it with our HS - operatorial  formulation, recapitulated in the previous section in the Euclidean case and introduce the basic tools like spectral triples for both commutative and non-commutative spaces, apart from the algebraic version of causality required to facilitate a similar computation of distances in commutative and non-commutative (NC) Lorentzian plane to be taken up in the next section.   \\ \\
We begin by providing a brief  outline of the very basic ingredients required to construct a Lorentzian spectral triple. This is given by the set of data ($\mathcal{A},\mathcal{H},\mathcal{D},\mathcal{J},\chi$) with 
\begin{itemize}
\item{A Hilbert space $\mathcal{H}$}
\item{A non-unital algebra $\mathcal{A}$ with a suitable and faithful representation $\pi$  on $\mathcal{H}$}
\item{A  Dirac operator $\mathcal{D}$ taken to be an unbounded operator, but $[\mathcal{D},\pi(a)]$ is bounded.}
\item{An operator $\mathcal{J}$  called Fundamental symmetry acting on $\mathcal{H}$  satisfying boundedness condition along with   $\mathcal{J}^2=1,\, \mathcal{J}^*=\mathcal{J},\,[\mathcal{J},a]=0,\, \forall a\in \tilde{\mathcal{A}},  \,\, \textrm{where} \,\,\tilde{\mathcal{A}}\,\, \textrm{is a unitized version of} \, \mathcal{A}$ and the Dirac operator fulfills $ \mathcal{D}^*=-\mathcal{J}\mathcal{D}\mathcal{J}$}
\item{For an even Lorentzian spectral triple we take $\chi$ to be  a grading operator  fulfilling the following conditions: 
\begin{equation}
\chi^*=\chi;\, \chi^2=1;\,\,\{\chi,\mathcal{J}\}=0;\,\{\mathcal{D},\chi\}=0
\end{equation}}
\end{itemize}
 The operator $\mathcal{J}$ captures the Lorentzian signature of the space by turning the Hilbert space into a Krein space where the positive definite inner product of the Hilbert space:  $\langle\cdot,\cdot\rangle$ turns to an indefinite inner product  space :$\,\,(\cdot,\cdot) =\langle\cdot,\mathcal{J}\cdot\rangle$ upon the insertion of the fundamental symmetry operator $\mathcal{J}$, so the Dirac operator in this inner product space becomes a non-hermitian operator and to render it as a skew Krein self-adjoint operator,  the condition $\mathcal{D}^*=-\mathcal{J}\mathcal{D}\mathcal{J}$ is imposed. We shall elaborate on this construction in section  4.
 % CAusal Cone %%%%%%%%%%%%%%%%%%% 
 
 \subsubsection{ Commutative Lorentzian Spectral Triple and Distance Formula}
 We now discuss, in this subsection, the example of a  commutative Lorentzian space-time taken to be globally hyperbolic  $n$ dimensional manifold $\mathcal{M}$  which is described by a  commutative spectral triple \cite{Franco1} constructed with :
 \begin{itemize}
 \item{Hilbert space $\mathcal{H}=L^2(\mathcal{M},S)$ of square integrable spinorial sections over $\mathcal{M}$}
 \item{Dirac operator $\mathcal{D}=-i\gamma^{\mu}\nabla_{\mu}$}
 \item{Algebra  $\mathcal{A}=C^{\infty}_0(\mathcal{M})$ of infinitely differentiable smooth real functions  (Ideally , one should introduce the unitized version of the algebra- a fact that we are ignoring for the time being. We shall, however, will make some pertinent observations later in the paper. )}
 \item{Fundamental symmetry $\mathcal{J}= i\gamma^0$ where $\gamma^0 $ is the time- component of Dirac's $\gamma$- matrices $\gamma^{\mu}$ satisfying the fundamental Clifford algebra : $\{\gamma^{\mu},\gamma^{\nu}\}=2\eta^{\mu\nu}\textbf{I}$ and the flat metric $\eta_{\mu\nu}=$diag(-,+,+,+)  }
 \item{Grading operator $\chi=(-i)^{\frac{n}{2}+1}\gamma^0\cdots\gamma^{n-1}$}
\end{itemize}  

For an even dimensional Lorentzian manifold i.e. if $n$ is even, then the distance between two points is essentially identified with the distance between the associated  pure states given by Dirac's $\delta$- functional serving as evaluation maps and is  given by \cite{Franco1}
\begin{equation}
d(p,q)=\inf_{a\in\mathcal{B}}\{[\delta_q(a)-\delta_p(a)]^+\}=\inf_{a\in \mathcal{B}}\{[a(q)-a(p)]^+ \}\label{f}
\end{equation} 
where $[\alpha]^+=\textrm{max}\{0,\alpha\},$ and
\begin{equation}\mathcal{B}=\{a\in\mathcal{A} :\,\textbf{B}:= \langle\phi,\mathcal{J}([\mathcal{D},a]\pm i\chi)\phi\rangle< 0,\,\,\,\forall\phi\in\mathcal{H}\}\subset \mathcal{A}\label{ballcond}
\end{equation}
  In the next subsection, following \cite{Franco1,Eckstein}, we shall see how this particular Ball condition essentially captures the conditions of  steep functions as a subset of causal functions in an algebraic set-up. 
 
\subsubsection{Non-commutative Lorentzian Spectral Triple and Distance Formula}

Now the Spectral triple for Non-commutative Lorentzian space-time, taken to be 2D Moyal plane,  can easily  be generalised. It  is comprised  of 
\begin{itemize}
\item{a Hilbert space $\mathcal{H}=\mathcal{H}_c\otimes\mathbb{C}^2$}
\item{Non-commutative algebra $\mathcal{A}=\mathcal{H}_q$ with a suitable representation $\pi$, again taken to be diagonal one i.e. $\pi(a):=\textrm{diag}(a,a)$.}
\item{Dirac operator $\mathcal{D}=-i\gamma^{\mu}\nabla_{\mu}$}
\end{itemize}
The fundamental symmetry operator and grading operator are same as defined in the commutative spectral triple. For non-commutative plane, however,  the concept of  a point is absent . So  the distance will be calculated between two pure states $\rho_1$ and $\rho_2$, just  like its Euclidean counter-part in section 2. The  Lorentzian distance between two such states $\rho_1$ and $\rho_2$ is  defined as 
\begin{equation}
d(\rho_1,\rho_2)=\inf_{a\in\mathcal{B}}\{[\rho_2(a) - \rho_1(a)]^+\}\label{ld}
\end{equation} where  $[\alpha]^+ =\textrm{max}\{0,\alpha\}$ and the ball $\mathcal{B}$ is defined as  
 \begin{equation}
\mathcal{B}=\{a\,:\,\textbf{B}=\langle\phi,\mathcal{J}([\mathcal{D},\pi(a)]\pm i\chi)\phi\rangle < 0, \forall  \phi\in\mathcal{H} \}\subset\mathcal{A}\label{ball}
\end{equation}

\subsection{An Algebraic Construction of Causality}
In this section we provide a very  brief review following \cite{Franco1,Franco4,Eckstein}  to show how the effect of causality is captured as an algebraic condition  in Lorentzian space. \\
\textbf{Theorem :} A function $a\in C^1(\mathcal{M},\mathbb{R})$ is causal \textit{iff} 
\begin{equation}
\forall \phi\in\mathcal{H} ,\,\,\langle\phi,\mathcal{J}[\mathcal{D},a]\phi\rangle <  0 \label{causal}
\end{equation}
where $\mathcal{J},\mathcal{D},\mathcal{H}$ are   defined  as above.  A sketch of the proof of this algebraic version can be given as following :\\
Recall that the causal property of  an absolutely continuous real valued $C^1$ function $a\in C^1(\mathcal{M},\mathbb{R})$,  can be fully characterised by  two conditions i.e.
\begin{equation}
\eta(\nabla a,\nabla a)< 0 \Rightarrow a_{,i}^2< a_{,0}^2 \,\,\,\,\,\,\,\textrm{and}\,\,\,\,\,\,\,\, \eta(\nabla a, \nabla T)=-a_{,0}< 0 \label{conditionnn}
\end{equation}
where $T$  is taken to be a temporal function and is typically taken to be $T=x^0 $  itself so that $\nabla T$ is  a time like vector and we have used a $n$-dimensional Lorentzian  flat metric  $\eta_{\mu\nu}=\textrm{diag}(-,+,+,+)$. If the conditions (\ref{conditionnn}) is false at some point of the manifold then from continuity of derivative it will be false in some of its neighbourhood as well.  Now  using $\mathcal{J}=i\gamma^0$  and $\mathcal{D} = -i\gamma^{\mu}\partial_{\mu}$, as above , we readily obtain, 
 \begin{equation}
 \mathcal{J}[\mathcal{D},a]= -a_{,0}+K;\,\,\,\,\,\,\, K:= \gamma^0\gamma^ia_{, i}
 \end{equation}
 Now this  $K$ can be shown to satisfy $K^2=\eta^{ij} \partial_i a\partial_j a $, so that the spectrum of  $K$ is essentially given by Spec$K= \{\pm\Vert\eta^{ij}\partial_i a\partial_j a\Vert^{\frac{1}{2}}\}$. Now since the reduced metric $\eta^{ij}$ is positive definite everywhere (in fact , in our case it is just identity matrix),  the matrix  $\mathcal{J}[\mathcal{D},a ] $ is point-wise negative and one readily verifies (\ref{causal}).\\ \\ \\
An important subspace of the causal function , is given by so called \textit{steep functions} , fulfilling  the property,
\begin{equation}
\eta(\nabla a,\nabla a) < -1.
\end{equation}
 A corresponding algebraic version of causality of this steep functions can be shown to be  given by\footnote{Note that the equality signs in the inequalities (\ref{causal},\ref{ccausal}) have been omitted here.This is  in contrast to \cite{Franco2,Franco4,Eckstein}. The  reason for this is  explained in the sequel (see section 4.1).}
\begin{equation}
\forall\phi\in\mathcal{H}, \langle\phi,\mathcal{J}([\mathcal{D},a]\pm i\chi)\phi\rangle < 0\label{ccausal}
\end{equation} 
An elementary proof of this algebraic version has been provided in \cite{Franco1,Franco4} , for even dimensional flat Lorentzian manifold. The basic idea behind the proof is to consider, the product manifold $\tilde{\mathcal{M}}=\mathcal{M}\times \mathbb{R}$ with the associated metric 
\begin{equation}
\tilde{\eta}= \left( \begin{array}{c|c}
\eta & 0 \\
\hline 
0&1
\end{array} \right)
\end{equation}
and introduce the  function $\tilde{a}:= a-x^n \in C^1(\mathcal{\tilde{M}},\mathbb{R})$. One can then show that $\tilde{a}$ is causal in $(\mathcal{\tilde{M}},\tilde{\eta})$ implies that $a$ is a steep function in $(\mathcal{M},\mathbb{R})$. \\The stage is therefore ready for its generalisation  to NC spaces.  All that we need to do there is to introduce an appropriate NC algebra $\mathcal{A}$ and replace the above real function $a$  by  suitable hermitian elements  $a\in \mathcal{\tilde{A}}$ (where $\tilde{\mathcal{A}}$ is the unitized version of $\mathcal{A}$), fulfilling
\begin{equation}
\forall\phi\in\mathcal{H}, \langle\phi,\mathcal{J}([\mathcal{D},\pi (\hat{a})]\pm i\chi)\phi\rangle < 0 \label{equ1}
\end{equation} 
 Having in our hand, an algebraic condition (\ref{equ1}) to capture the space of steep functions  on the manifold we shall  use it  as  \textit{Ball condition} in Lorentzian manifold.  \\ \\ 
 Now we can introduce a convex cone $\mathcal{C}$ to be the set of  causal algebra elements  satisfying (\ref{causal}) fulfilling a further condition ,
\begin{equation}  \overline{Span_{\mathbb{C}}\mathcal{C}} = \overline{\tilde{\mathcal{A}}}\label{causal2}
\end{equation} 
By this a  partial order relation is induced  on the space of states $S(\mathcal{A})$, which is called the causal relation \cite{Wallet},
\begin{equation}
\forall \omega_1,\omega_2 \in S(\mathcal{A}), \omega_1 \preceq\omega_2 \,\,\,\textrm{iff}\,\,\, \forall a \in \mathcal{C}, \omega_1(a) \leq \omega_2(a)
\end{equation} 

 \section{Construction of Dirac Operator,  Krein Space and Ball Condition}
In this section we try to mimic our Hilbert-Schmidt formulation as described in section 2 to calculate the distances in Lorentzian commutative and non-commutative spaces. Both of the spaces have the  same structure of  Dirac operator $\mathcal{D}=\hat{P}_{\mu}\otimes\gamma^{\mu}$, which by default act on the Hilbert spaces $L^2(\mathbb{R}^{1,1})\otimes\mathbb{C}^2$ and  $\mathcal{H}_q\otimes\mathbb{C}^2$ respectively for the commutative and non-commutative planes. A  generic spinor  of this Hilbert space can be written as  $\Psi= \begin{pmatrix}
|\psi_1) \\|\psi_2) 
\end{pmatrix}$ ,with $|\psi) \in L^2(\mathbb{R}^{1,1})$ and $\mathcal{H}_q $  in the respective cases.\\ \\
For pseudo-euclidean signature (-,+)   of  our  space-time  $\mathbb{R}^{1,1}$,  we can take $\gamma^{0}=-i\sigma^1; \gamma^1=\sigma^2$ so that  $(\gamma^0)^2=-1$ , $(\gamma^1)^2=1$ and $ \{\gamma^{\mu},\gamma^{\nu}\}=2\eta^{\mu\nu}$. With these the Dirac operator takes the following form:

% ^^^^^^^^^^^^^^^^^^^^^^DIRAC  OPERATOR^^^^^^^^^^^^^^^^^^^^^^
\begin{equation}
\mathcal{D} =-i\begin{pmatrix}
0&  \hat{P}_0+\hat{P}_1\\\hat{ P}_0- \hat{P}_1&0
\end{pmatrix}
\label{d}
\end{equation}
%^^^^^^^^^^^^^^^^^^^^^^^^^^^^^^^^^^^^^^^^^^^^^^^^^^^^^^^^^^^^^^^^^^
 %where we have represented $\hat{P}^0$ and $\hat{P}^1$ by $i\partial_t$  and  $-i\partial_x$ respectively. 
 As it stands,  the Dirac operator is not a self adjoint operator. We can however  convert the Hilbert space into a Krein space ,  where the operator will become  a skew Krein self-adjoint operator.  It will therefore be useful to recall the definition of Krein space first and study a simple example of $\mathbb{C}^2$ , whose Krein counter-part will be constructed and eventually be used to construct the Krein counterpart of the entire Hilbert space $\mathcal{H}=\mathcal{H}_q\otimes \mathbb{C}^2$ occurring in the above spectral triple.\\
 \\ \textbf{Def}:\\  \vspace{3pt} If an indefinite inner product  space $\mathcal{K}$  can be split into two sub-spaces as $\mathcal{K}=\mathcal{K}_+\oplus \mathcal{K}_-$ which are mutually orthogonal,  complete (in the norm induced on them) having positive and negative definite inner product respectively and if  further the inner product on $\mathcal{K}$ is  non degenerate, then $\mathcal{K}$ is called a Krein space \cite{Franco2,Franco4}.\\  \\For every such decomposition we can associate a fundamental symmetry operator $\mathcal{J}$ as $\mathcal{J}=\textrm{id}_+ \oplus (-\textrm{id}_-)$ respecting $\mathcal{J}^2=1$ and whose insertion can turn the  subspace  with negative inner product $(\cdot,\cdot)$ to the sub-space with a positive definite inner product $\langle\cdot,\cdot\rangle$  and vice-verse with the following operation : $ (\cdot, \mathcal{J}\cdot)=\langle \cdot ,\cdot\rangle$ or equivalently $\langle\cdot,\mathcal{J}\cdot\rangle = (\cdot,\cdot)$. \\ \\ \\
 \textbf{Converting $\mathbb{C}^2$ and $L^2(\mathbb{R}^{1,1})\otimes\mathbb{C}^2$ and $\mathcal{H}_q\otimes\mathbb{C}^2$ into Krein spaces} \\  \\
The 2 dimensional complex vector space $\mathbb{C}^2$ can  be regarded as direct sum of two sub-spaces $\mathbb{C}^1_{\pm}$ as follows ,
$$\mathbb{C}^2= \textrm{Span} \left\{\frac{1}{\sqrt{2}}\begin{pmatrix}
1\\1
\end{pmatrix},\frac{1}{\sqrt{2}}\begin{pmatrix}
1\\-1
\end{pmatrix}\right\} =\textrm{Span} \left\{\frac{1}{\sqrt{2}}\begin{pmatrix}
1\\1
\end{pmatrix}\right\}  \oplus \textrm{Span} \left\{\frac{1}{\sqrt{2}}\begin{pmatrix}
1\\-1
\end{pmatrix}\right\}:=\mathbb{C}^1_+\oplus \mathbb{C}^1_- $$
with associated projectors   $\mathbb{P}^{\pm} = \frac{1}{2}(\textbf{1}\pm \sigma_1)$  projecting  a generic two component vector $\begin{pmatrix}
\alpha\\ \beta
\end{pmatrix} \in \mathbb{C}^2 $ to  the  subspaces  $\mathbb{C}^1_+ $and  $\mathbb{C}_-^1$ respectively where both the subspaces are naturally endowed with positive definite inner products. Now let us take the fundamental symmetry operator $\mathcal{J}=\sigma_1$ which makes the $\mathbb{C}^1_-$ subspace a negative definite inner product space but retains the positive-definiteness property of  $\mathbb{C}^1_+$ space  with the following insertion : \begin{equation}
\langle\cdot,\mathcal{J}\cdot\rangle=\langle\cdot,\sigma_1\cdot\rangle =(\cdot,\cdot).\label{inn}
\end{equation} So now with the new inner product $(\cdot,\cdot)$  defined in (\ref{inn}) we can call our $\mathbb{C}^2$ space as a Krein space. \\ \\
With this it can now be shown trivially that  $\mathcal{H}_q\otimes \mathbb{C}^2$ equipped with the inner product $\langle\cdot,\mathcal{J}\cdot\rangle $,  where the used  fundamental symmetry operator now is $\mathcal{J}=\textbf{I}\otimes\sigma_1$ ,is an indefinite inner product space or Krein space where $\mathcal{H}_q\otimes\mathbb{C}^2$ splits as
\begin{equation*}
\mathcal{H}_q\otimes\mathbb{C}^2=( \mathcal{H}_q\otimes\mathbb{C}^1_+)\oplus(\mathcal{H}_q\otimes\mathbb{C}^1_-).
\end{equation*}  With this fundamental symmetry $\mathcal{J}$,  it can indeed be checked quite easily that  $\mathcal{D}^*=-\mathcal{J}\mathcal{D}\mathcal{J}$ proving $\mathcal{D}$  (\ref{d}) to be skew Krein self-adjoint,\\ \\
The grading operator as defined in section - 3.0.1 comes out to be just $\chi=\textbf{1}\otimes\sigma^3 $. We take a suitable representation of the algebra element $a$ as $\pi(a)=\textrm{diag}(a,a)$. Upto this all  the elements of spectral triples are assumed  to be quite  generic and applicable for both commutative and non-commutative Lorentzian plane. So a general Ball condition can  now be derived from  (\ref{ballcond}) as,
 %^^^^^^^^^^^^^^^^^^^^ BALL CONDITION^^^^^^^^^^^^^^^^^^^^^^
 
\begin{equation}
\textbf{B}=\langle\Psi, \mathcal{J}\{[\mathcal{D},\pi(a)]+i\chi\}\Psi\rangle =\begin{pmatrix}
\langle\psi_1|&\langle\psi_2|\end{pmatrix}
\begin{pmatrix}
-i[\hat{P_0}-\hat{P_1},a]&-i\\ i&-i[\hat{P_0}+\hat{P_1},a]
\end{pmatrix}
\begin{pmatrix}
|\psi_1\rangle\\|\psi_2\rangle
\end{pmatrix}< 0\,\,\, \forall\,\,\, \Psi\in\mathcal{H}\label{pbp}
\end{equation}
It can be re-cast as,
\begin{equation}
\textbf{B}= -i\langle\psi_1|[\hat{P}_0-\hat{P_1},a]|\psi_1\rangle - i\langle\psi_2|[\hat{P}_0+\hat{P}_1,a]|\psi_2\rangle+i[\langle\psi_2|\psi_1\rangle -\langle\psi_1|\psi_2\rangle ] <0 \label{ball2}
\end{equation}
Above equation can be taken as a master equation or  principal  ball condition. We now specialise into  two separate cases for commutative and non-commutative planes.

\subsection{Distance in Commutative Lorentzian Plane}
 As a warm up exercise we first undertake the computation  of  the distance on the (1+1) dimensional flat commutative Lorentzian manifold itself.  The spectral triple for commutative Lorentzian manifold $\mathbb{R}^{1,1}$ is defined through the algebra of smooth functions over $\mathbb{R}^{1,1}:\mathcal{A}=C^{\infty}_0(\mathbb{R}^{1,1})$, a Hilbert space \footnote{Here the $L^2(\mathbb{R}^{1,1})$ part of $\mathcal{H}$ is actually analogous to  $\mathcal{H}_q$
 in our HS formulation.} $\mathcal{H}=L^2(\mathbb{R}^{1,1}) \otimes \mathbb{C}^2$, consisting of  generic spinor like  $
\Psi= \begin{pmatrix}
|\psi_1) \\|\psi_2) 
\end{pmatrix}$ where $|\psi_i)$ are abstract kets whose  representation in $|t,x\rangle$ basis   becomes $L^2(\mathbb{R}^{1,1})$ element. With the Dirac operator  already defined in previous section we arrived at a simplified  ball   condition  (\ref{ball2}). Inserting  completeness relation of  $|t,x\rangle \equiv |x\rangle$ basis: $\int dt\,dx\,|t,x\rangle\langle t,x| =\textbf{1}$,  (\ref{ball2}) can be rewritten as,
\begin{align}
-i\int d^2x\,d^2y\, \psi_1^*(x)\psi_1(y)\langle x|[\hat{P}_0-\hat{P}_1,a]|y\rangle -i\int d^2x\, d^2y\,\psi_2^*(x)\psi_2(y)& \langle x|[\hat{P}_0+\hat{P}_1,a]|y\rangle \nonumber\\
&+i\int d^2x [\psi_2^*(x)\psi_1(x)-\psi_1^*(x)\psi_2(x)]<0\label{ball3}
 \end{align}
Here the action of  momentum operator on the  $C^{\infty}$ function $a(t,x):=a(x)$  takes place through commutator bracket and it  gives,
\begin{equation}
\langle x |[\hat{P}_{\mu},a(x)]|y\rangle = -i\partial_{\mu}a(x) \delta^{(2)} (x-y)  \label{hj}
\end{equation} 
With this   (\ref{ball3}) further simplies as
\begin{equation}
\int d^2x \,\,[ (\partial_0 a)(-|\psi_1(x)|^2-|\psi_2(x)|^2)+(\partial_1 a)(|\psi_1(x)|^2-|\psi_2(x)|^2)] < i\int d^2x\, [\psi_1^*(x)\psi_2(x)-\psi_2^*(x)\psi_1(x)] \label{hjj}
\end{equation}
Now this  condition is valid  for each and every point of the manifold for arbitrary $\psi_1(x)$ and $\psi_2(x)$. So this condition applies for the integrand itself point-wise and enables us to write, 
\begin{equation}
 (\nabla_{\mu}a)V^{\mu} < i(\psi_1^*\psi_2-\psi_2^*\psi_1) \,\,; V_0=-V^0=|\psi_1|^2+|\psi_2|^2, V_1= V^1=|\psi_1|^2-|\psi_2|^2 \label{first}
 \end{equation} 
 
 where $V=(V^0,V^1)$ is a time-like two vector  and $a\in \mathcal{A}$  belong to  the set of steep functions i.e.  $\eta(\nabla a,\nabla a) < -1 $ which means that $\nabla a$ too is a time-like vector and $(V^{\mu}\partial_{\mu}a)$ is intrinsically negative\footnote{Note that if equality signs were to be included in (\ref{ccausal}), $(V^{\mu}\nabla_{\mu}a)$ would not have been intrinsically negative and we would  be forced to consider the case $V^{\mu}\nabla_{\mu}a  =0$ also. For $V^{\mu}$ time-like, this would have implied that $\nabla_{\mu}a$ space-like - a scenario which we would like to avoid. In any case, the inclusion/exclusion does not impact the computation of supremum/infimum, as one is a dense subspace of the other.}, enabling us to recast (\ref{first}) as   \begin{equation}
 \vert V^{\mu}\partial_{\mu}a\vert = -(V^{\mu}\partial_{\mu}a)> i(\psi_1\psi_2^*-\psi_2\psi_1^*)= - 2|\psi_1||\psi_2|\sin(\alpha_1-\alpha_2)\label{eq3}
 \end{equation}
 where the phases $\alpha_1$ and $\alpha_2$ are defined as  $\psi_1= |\psi_1|e^{i\alpha_1}$ and $\psi_2=|\psi_2|e^{i\alpha_2}$. Now since the computation of  Lorentzian distance in (\ref{f}) involves the computation of  infimum we have to search for the optimal algebra element for which the infimum is attained . For that, it will be advantageous to consider a slight variant of the inequality (\ref{ball3},\ref{eq3}) as
 \begin{equation}
 \textbf{B}\leq 0\,\,  \textrm{i.e.}\,\, \vert V^{\mu}\partial_{\mu}a\vert \geq -2|\psi_1||\psi_2|\sin(\alpha_1-\alpha_2)\label{A1}
\end{equation} 
The inclusion of the equality sign here will  not have any effect on the computation of the infimum, as the set determined by (\ref{ball3},\ref{eq3}) is a dense subset of (\ref{A1}). At this stage we we first hold $|\psi_1|$ and $|\psi_2|$ fixed and vary $\alpha_1,\alpha_2$ so that the saturation condition in (\ref{A1}) is satisfied. Clearly the maximal value reached by the R.H.S above  will correspond to $ 2|\psi_1||\psi_2|$ with the choice $\alpha_2-\alpha_1=\frac{\pi}{2}$ and (\ref{A1}) reduces to the following form 
\begin{equation}
\vert V^{\mu}\partial_{\mu}a\vert\geq 2|\psi_1||\psi_2|\label{A2}
\end{equation}
 Correspondingly the generic form of the spinor $\psi$ which maximises the R.H.S is $\begin{pmatrix}
|\psi_1|\\i|\psi_2|
\end{pmatrix}e^{i\alpha_1}$. 
On the other hand we can now apply \textit{reverse} Cauchy-Schwarz  inequality \cite{Franco1} to write 
\begin{equation}
\vert\partial_{\mu}a V^{\mu}\vert \geq \Vert\nabla_{\mu}a\Vert_L \Vert V^{\mu}\Vert_L. \label{eq9}
\end{equation}
Note that here L in the subscript is a reminder of Lorentzian norm  defined as $\Vert v\Vert _L= \sqrt{-\eta(v,v)}$ for a time-like vector $v$. In the next stage , we vary $|\psi_1|$ and $|\psi_2|$, so that  $V$ becomes collinear with  $\nabla a$  and (\ref{eq9}) becomes, 
\begin{equation}
\vert V^{\mu}\partial_{\mu}a \vert = \Vert\nabla_{\mu}a\Vert_L \Vert V^{\mu}\Vert_L\label{eq2}
\end{equation}  
We now make use of (\ref{first}), so that by (\ref{A1}) and (\ref{eq2}) we finally get 
\begin{equation}
\Vert \nabla a\Vert_L \geq 1\label{eq1}
\end{equation}
 We therefore look for a solution of $a$ satisfying  
\begin{equation}
\Vert\nabla a \Vert_L = 1\label{eq12}
\end{equation}
Equivalently , the  search can be restricted to the solution set of the following differential equation satisfied by  real functions $a(x,t)$ satisfying  
\begin{equation}
\left(\frac{\partial a}{\partial t}\right)^2-\left(\frac{\partial a}{\partial x}\right)^2=1\label{eq4}
\end{equation}
yielding the following one-parameter $(\lambda)$  solution set of $a(t,x)$:
%We parametrize the coordinates as points in belonging to forward light-cone as $t=r\cosh \phi;\,\,\, x= r\sinh\phi$ and rewrite  (\ref{eq4}) as 
%\begin{equation}
%\left(\frac{\partial a}{\partial t}\right)^2-\frac{1}{r^2}\left(\frac{\partial a}{\partial \phi}\right)^2=1\label{eq5}
%\end{equation}
\begin{equation}a(t,x)=t\cosh\lambda +x\sinh\lambda\label{form} 
\end{equation} 

By (\ref{f}) the distance between two points say   $P(t_1,x_1),Q(t_2,x_2)$ in commutative Lorentzian manifold (where $P$ preceeds $Q$ chronologically  i.e. P $\prec$ Q) is now given by, 
\begin{equation}
d(\delta_{P},\delta_{Q})=\inf_{a\in\mathcal{B}}\{[\delta_{Q}(a)-\delta_{P}(a)]^+\} =\inf_{a\in\mathcal{B}}\{[a(t_2,x_2)-a(t_1,x_1)]^+\}\label{m}
\end{equation}
where $\delta_P,\delta_Q$ are pure states representing  the respective events $(t_1,x_1)$ and $(t_2,x_2)$ in the forward light-cone of the commutative plane. Eq (\ref{m}) can now be recast as,
\begin{equation}
d(\delta_{P},\delta_{Q})= \inf_{\lambda}\{[(t_2-t_1)\cosh\lambda+(x_2-x_1)\sinh\lambda]^+\}\label{dis1}
\end{equation} 
which on minimisation with respect to $\lambda$ (to get the infimum value of the set ) gives ,
\begin{equation}
d(\delta_{P},\delta_{Q})= \sqrt{(t_2-t_1)^2-(x_2-x_1)^2}\label{dis}
\end{equation}
Note the resultant distance in (\ref{dis1}) automatically contains the causal information.  If the events are not causally connected i.e.if $Q$ is not preceeded by $P$ then  the quantity in the parenthesis of (\ref{dis1}) will definitely be negative \footnote{For $Q\prec P, \,\, t_1>t_2$ and  $ (t_2-t_1)^2>(x_2-x_1)^2$ . Also $ |\cosh\lambda| >|\sinh\lambda|\,\,\, \forall\lambda$. Consequently whole quantity in (\ref{dis1}) will always be negative.} and the distance formula  will result in zero value indicating that the distance is not symmetric under the interchange of  pair of events.\\

\subsubsection{Functional Derivative  Approach}
 In this  sub-section we shall try to address the same problem with functional differentiation approach which we will  find to be easier to execute and will be easily adaptable to the case of computation of   the distance in Lorentzian Moyal plane to be taken up  in next sub-section. For this let us consider  the left hand side of the ball condition (\ref{ball2}) , $\textbf{B}[\psi_1,\psi_1^*,\psi_2,\psi_2^*]$ which can now be regarded as a functional of $\psi_1,\psi_1^*,\psi_2,\psi_2^*$:
 \begin{equation}
\textbf{B}[\psi_1,\psi_1^*,\psi_2,\psi_2^*]= -\int d^2x \left[ |\psi_1(x)|^2(\partial_0 a-\partial_1 a)+|\psi_2(x)|^2(\partial_0 a+\partial_1 a)+i\{\psi_1^*(x)\psi_2(x)-\psi_2^*(x)\psi_1(x)\}\right]  \label{hjp}
\end{equation}
Clearly the algebra element $a\in\mathcal{A}$  for which the infimum in (\ref{f})  is reached giving the distance should be such that  
\begin{equation}
\sup_{\psi_1,\psi_2,\psi_1^*,\psi_2^*}\,\,\, \textbf{B}[\psi_1,\psi_2,\psi_1^*,\psi_2^*]=0\label{supremum}
\end{equation}
In other words we want to extract a condition on the algebra element for which the functional $\textbf{B}[\psi_1,\psi_1^*,\psi_2,\psi_2^*]$ reaches its supremum value of zero. All that  we need to do first is to maximise \textbf{B} when $\psi_1, \psi_2, \psi_1^*, \psi_2^*$ are varied arbitrarily, and show  that the maximum of $\textbf{B}[\psi_1,\psi_1^*,\psi_2,\psi_2^*]$ corresponds to the  zero value which coincides with the supremum, only with a suitable constraint on the algebra element  $a\in\mathcal{A}$, which emerges  as an offshoot. When  the functional  $\textbf{B}[\psi_1,\psi_1^*,\psi_2,\psi_2^*]$ is functionally differentiated with respect to $\psi_1(x),\psi_2(x) $, we get the following matrix equation,
\begin{equation}
\begin{pmatrix}
-(\partial_0-\partial_1)a(x)&i\\
-i&-(\partial_0+\partial_1)a(x)
\end{pmatrix}
\begin{pmatrix}
\psi_1^*(x)\\ \psi_2^*(x)
\end{pmatrix}=0\label{matrix}
\end{equation} 
The variation with respect to $\psi_1^*(x)$ and $\psi_2^*(x)$ yields an equivalent matrix equation, which is just  the complex conjugate of the above equation. If  $\psi_1^*,\psi_2^*$ are zero then the equations are trivially satisfied. For allowing non-vanishing values of $\psi_1^*$ and $\psi_2^*$, albeit a linear and consistent pair of equations relating them, we must have vanishing determinant of the coefficient matrix  in (\ref{matrix}) yielding a condition on $a(x)$, which is precisely (\ref{eq4}) itself, giving (\ref{form}) for $a(t,x)$ . 
Now let us rewrite \textbf{B}  using (\ref{form}) as,
\begin{equation*}
\textbf{B}=-\int d^2x\,[e^{-\lambda}\psi_1^*(x)\psi_1(x)+e^{\lambda}\psi_2^*(x)\psi_2(x)+i\{\psi_1^*(x)\psi_2(x)-\psi_2^*(x)\psi_1(x)\}]
\end{equation*}
The following scaling transformations, $\psi_1(x)\to \psi_1^{\prime}(x):= e^{-\frac{\lambda}{2}}\psi_1(x);\,\, \psi_2(x)\to \psi_2^{\prime}(x):= e^{\frac{\lambda}{2}}\psi_2(x)$ allows us to write \textbf{B} in a simplified form as
\begin{equation}
\textbf{B}=-\int d^2x\,[\psi_1^{\prime *}(x)\psi_1^{\prime}(x)+\psi_2^{\prime *}(x)\psi_2^{\prime}(x)+i\{\psi_1^{\prime *}(x)\psi_2^{\prime}(x)-\psi_2^{\prime *}(x)\psi_1^{\prime}(x)\}]
\end{equation} 
We now  carry out a linear transformation of the following form which enables us to write \textbf{B} in terms of  independent functions $\phi_{\pm}(x)$, defined as,
\begin{equation}\begin{pmatrix}
\phi_+(x)\\ \phi_-(x)
\end{pmatrix}=\frac{1}{2}
\begin{pmatrix}
1&i\\-1&i
\end{pmatrix}
\begin{pmatrix}
\psi_1^{\prime }(x)\\ \psi_2^{\prime }(x)
\end{pmatrix}\label{trans5}
\end{equation}
A simple algebra then shows that \textbf{B}  is really independent of $\phi_{+}(x)$ and depends on $\phi_-(x)$ only as,
\begin{equation}
\textbf{B}=-2\int d^2x |\phi_-(x)|^2 \label{max}
\end{equation}
Equation (\ref{max}) represents  an inverted parabola  in ($\textbf{B}, |\phi_-|$) plane  and clearly implies that , \textbf{B} reaches its supremum value zero for $|\phi_-|=0$. So the form of  the algebra element   (\ref{form})  automatically ensures that \textbf{B} can not exceed zero whatever be $\phi_+$ and $\phi_-$. We therefore arrive  eventually at the same distance function  (\ref{dis}).
 \subsection{Distance in 2D  Lorentzian Moyal Plane}
 We finally take up the computation of the distance in the non-commutative  case in this pen-ultimate section. To begin with note that a representation of the non-commutative 2-dimensional  Moyal plane algebra : $[\hat{t},\hat{x}]=i\theta$   with Lorentzian signature (-,+) will be  furnished again by  the same Hilbert space $\mathcal{H}_c$ defined in (\ref{k}). Here  the creation and annihilation operators along with the \textit{vacuum} $|0\rangle$ are defined analogously as : 
 \begin{equation}
 \hat{b}=\frac{\hat{t}+i\hat{x}}{\sqrt{2\theta}},\,\,\hat{b}^{\dagger}=\frac{\hat{t}-i\hat{x}}{\sqrt{2\theta}};\,\,\,\hat{b}|0\rangle =0 \label{b5}
  \end{equation}
However there are certain differences as well. To see this note that although the vacuum transforms under space-time translation as,\begin{equation}|0\rangle \rightarrow |z\rangle =U(z,\bar{z})|0\rangle;\,\,\,U(z,\bar{z})= e^{-\bar{z}\hat{b}+z\hat{b}^{\dagger}}\label{coh2}
\end{equation}
along all the raising/lowering operators  transforming adjointly : 
\begin{align}
&\hat{b}\rightarrow\hat{b}_z=\hat{b}-z=U(z,\bar{z})\hat{b}U^{\dagger}(z,\bar{z});\,\,\,\hat{b}^{\dagger}\rightarrow\hat{b}_z^{\dagger}=\hat{b}^{\dagger}-\bar{z}=U(z,\bar{z})\hat{b}^{\dagger}U^{\dagger}(z,\bar{z});\nonumber\\
&\rho_0\rightarrow\rho_z=|z\rangle \langle z| =U \rho_0 U^{\dagger} ;\,\,\mathcal{D}\rightarrow\mathcal{D}_z= U(z,\bar{z})\mathcal{D}U^{\dagger}(z,\bar{z})\label{trans}
 \end{align}
 with $U(z,\bar{z})$ (\ref{coh2}) being the same as its Euclidean counter-part, the situation is different when the Lorentz - boost is concerned. In fact the transformation under Lorentz boost, 
 \begin{equation}
\begin{pmatrix}\hat{t}\\\hat{x}\end{pmatrix} \rightarrow \begin{pmatrix} \hat{t}^{\prime}\\\hat{x}^{\prime}\end{pmatrix} = \begin{pmatrix} \cosh\phi&\sinh\phi\\ \sinh\phi&\cosh\phi\end{pmatrix}\begin{pmatrix}\hat{t}\\ \hat{x}\end{pmatrix}\label{E1}
  \end{equation}, is unitarily implemented on Hilbert space $\mathcal{H}_c$ through adjoint action as:
\begin{align}
&\hat{b}\rightarrow\hat{b_{\phi}}=\cosh\phi\,\hat{b}+i\sinh\phi\,\hat{b}^{\dagger}= U(\phi)\hat{b}U^{\dagger}(\phi) ;\,\,\,\,\hat{b}^{\dagger}\rightarrow\hat{b_{\phi}}^{\dagger}=\cosh\phi\,\hat{b}^{\dagger}-i\sinh\phi\,\hat{b}=U(\phi)\hat{b}^{\dagger}U^{\dagger}(\phi)\nonumber\\
& \rho_0\rightarrow\rho_{\phi}=|\phi\rangle\langle\phi |=U(\phi)\rho_0 U^{\dagger}(\phi);\,\,\,\mathcal{D}\rightarrow\mathcal{D}_{\phi}=U(\phi)\mathcal{D}U^{\dagger}(\phi)\label{LT}
 \end{align}
where  $U(\phi)= e^{-\frac{i\phi}{2}(\hat{b}^{2}+{\hat{b^{\dagger}}}^2)}$ is now a squeezing operator. Note that  the coherent state also transforms by left action as  $|z\rangle \rightarrow |z;\phi\rangle = U(\phi)|z\rangle$, like all vectors $|\psi\rangle\in\mathcal{H}_c$ fulfilling $b_{\phi}|z;\phi\rangle =z|z;\phi\rangle $. In particular, the \textit{vacuum} state $|0\rangle\in\mathcal{H}_c$  which also belongs to the family of coherent states with $z=0$ is found to be non-invariant under this boost : $|0;\phi\rangle =U(\phi)|0\rangle \neq |0\rangle$  like its associated pure state $|0;\phi\rangle \langle 0;\phi| \neq  |0\rangle\langle 0|$ and this is in contrast to its counter-part in the Euclidean case where the vacuum $|0\rangle$ picks up a simple phase under SO(2) rotation so that the associated pure states $|0\rangle\langle 0|$ remains invariant. And this is despite the fact that the space-time algebra $[\hat{t},\hat{x}]=i\theta$ along with the ball condition is invariant under this Lorentz transformation also. Nevertheless, one can easily prove the following identity:
\begin{equation*}
\langle z ,\phi|\hat{a}_{w,\phi}|z,\phi\rangle - \langle w,\phi|\hat{a}_{w,\phi}|w,\phi\rangle = \langle z-w|\hat{a}|z-w\rangle -\langle 0|\hat{a}|0\rangle;\,\, \hat{a}_{z,\phi}:=U(\phi)\hat{a}_zU^{\dagger}(\phi)
\end{equation*}
so that the spectral distance is invariant under both translation and Lorentz boost:
\begin{equation}
d_{\mathcal{D}}(\rho_0,\rho_{z-w}) = d_{\mathcal{D}_{z,\phi}}(\rho_{w,\phi},\rho_{z,\phi})\label{ISO}
\end{equation}
proving the invariance of distance  under Poincare transformation.  In (1+1) dimension we have three generators e.g. two translations and one boost  for the transformation which forms a closed ISO(1,1)  algebra among themselves. \\ \\
We  finally embark on the explicit computation of the distance. For  this we shall essentially follow the same footsteps in HS formalism  as shown in section-2  for Euclidean Moyal plane . We construct the spectral triple $(\mathcal{A},\mathcal{H},\mathcal{D})$ here with  the algebra $\mathcal{A}=\mathcal{H}_q$, the    Hilbert space  $\mathcal{H}=\mathcal{H}_q\otimes \mathbb{C}^2$, and the Dirac operator $\mathcal{D}=\hat{P}_{\mu}\otimes \gamma^{\mu}$  (\ref{d}) which again by default  acts  on  $\mathcal{H}_q\otimes\mathbb{C}^2$.  Here the momentum operators have an adjoint action on an element $|\psi)\in\mathcal{H}_q$ as:
 \begin{equation}
 \hat{P}_i|\psi)=\frac{\epsilon_{ij}}{\theta}|[\hat{x}_j,\psi]); \,\, i,j\in[0,1]\label{adjoint}
\end{equation}  
The commutators occurring  in the ball condition (\ref{ball2})  now becomes,
\begin{equation}
[\hat{P}_0+\epsilon\hat{P}_1,\hat{a}]=\frac{1}{\theta}[\hat{x}-\epsilon\hat{t},\hat{a}];\,\,\,\epsilon=\pm 1,\,\,a\in\mathcal{A}=\mathcal{H}_q \label{adjoint2}
\end{equation} 
It is now evident that instead of $\mathcal{H}_q\otimes\mathbb{C}^2$ as our Hilbert space in the spectral triple , we can again take it to be  $\mathcal{H}_c\otimes\mathbb{C}^2$ as $\hat{x},\hat{t}$ has normal left action on $\mathcal{H}_c$, so that the matrix elements corresponding to the first two terms in (\ref{ball2}) are now well defined with $|\psi_1\rangle$ and $|\psi_2\rangle\in\mathcal{H}_c$. However in contrast to the commutative case, discussed earlier, we don't  have any concept of a point or rather a well defined event .The best option is to use coherent states (\ref{coh}) which enjoys the desirable feature that they are maximally localised space-time event:$\Delta\hat{t}\Delta\hat{x}=\frac{\theta}{2}$. Consequently , it will be advantageous again to take recourse to the Fock-Bergman representation (Appendix A) of coherent state basis. By inserting completeness relation (\ref{A68}) in  the Ball condition (\ref{ball2}) this  gives the functional,
\begin{align}
\textbf{B}[\psi_1,\psi_1^*,\psi_2,\psi_2^*]= -i\int d\mu (z,\bar{z})\, d\mu(w,\bar{w})\,[ \psi_1^*(z)\psi_1(\bar{w})\langle \bar{z}|& [\hat{P}_0-\hat{P}_1,a]|w\rangle +\psi_2^*(z)\psi_2(\bar{w}) \langle \bar{z}|[\hat{P}_0+\hat{P}_1,a]|w\rangle]\nonumber\\
&+i\int d\mu (z,\bar{z})\, [\psi_2^*(z)\psi_1(\bar{z}) -\psi_1^*(z)\psi_2(\bar{z})] <0 \label{ball4}
\end{align}
Using (\ref{adjoint2}) we have,
\begin{align}
\textbf{B}[\psi_1,\psi_1^*,\psi_2,\psi_2^*]=-\frac{i}{\theta}\int d\mu (z,\bar{z})\, d\mu(w,\bar{w})\,& [\psi_1^*(z)\psi_1(\bar{w})\langle \bar{z}| [\hat{t}+\hat{x},\hat{a}]|w\rangle -\psi_2^*(z)\psi_2(\bar{w}) \langle \bar{z}|[\hat{t}-\hat{x},\hat{a}]|w\rangle] \nonumber\\
&+i\int d\mu (z,\bar{z})\, [\psi_2^*(z)\psi_1(\bar{z})-\psi_1^*(z)\psi_2(\bar{z})] <0 \label{ball9}
\end{align}
We take the left hand side of (\ref{ball9}) as a functional $\textbf{B}[\psi_1,\psi_1^*,\psi_2,\psi_2^*]$ of $\psi_1(\bar{z}),\psi_1^*(z),\psi_2(\bar{z}),\psi_2^*(z)$. To evaluate the optimal algebra element which saturates the ball condition we maximise \textbf{B} with respect to $\psi_1(\bar{u})$ and $\psi_2(\bar{u})$ using (\ref{A71}) and equate them to zero,  to arrive respectively at the following pair of equations:
\begin{align}
\frac{1}{\theta}\langle\psi_1|[\hat{t}+\hat{x},\hat{a}]|u\rangle &=\langle\psi_2|u\rangle \nonumber\\
\frac{1}{\theta}\langle\psi_2|[\hat{t}-\hat{x},\hat{a}]|u\rangle &= \langle\psi_1|u\rangle\label{condition}
\end{align} 
As can be checked easily  that the differentiation w.r.t. $\psi_1^*(u)$ and $\psi_2^*(u)$ just yield an equivalent pair of equations related to  (\ref{condition}) by just a complex conjugation. Combining this pair of equations by using the completeness relation (\ref{A68}), one readily obtains a condition on the algebra element as,
\begin{equation}
\frac{1}{\theta^2}[\hat{t}+\hat{x},\hat{a}]\,[\hat{t}-\hat{x},\hat{a}]= 1\label{con3}
\end{equation} 
Like in section-2 here too the general form of the optimal algebra element $\hat{a}_s$ which is an infinite dimensional matrix satisfying  (\ref{con3}), which can also  ensure  ISO(1,1) invariance of the resulting distance will clearly depend linearly on $\hat{t}$ and $\hat{x}$ as: $\hat{a}=\alpha\hat{t}+\beta\hat{x}$(with $\alpha,\beta\in\mathbb{R}$ for $\hat{a}$ being hermitian) . Substituting this in (\ref{con3}) , we finally arrive at the general form of the algebra element: 
\begin{equation}
\hat{a}_{\lambda}=\hat{t} \cosh\lambda+\hat{x}\sinh\lambda;\,\,\,\,\lambda\in\mathbb{R}\label{aform2}
\end{equation}  
To see it more concretely, let us  we can write (\ref{con3}) in form of   differential equations.For that we consider the matrix elements of the above operator equation (\ref{con3}) in coherent state basis as,  
\begin{equation*}
\int\, d\mu(w,\bar{w})  \langle\bar{z}|[e^{\alpha}\hat{b}+e^{-\alpha}\hat{b}^{\dagger},\hat{a}]|w\rangle\langle\bar{w}|[e^{-\alpha}\hat{b}+e^{\alpha}\hat{b}^{\dagger},\hat{a}]|u\rangle = \theta \int \, d\mu(w,\bar{w}) \langle\bar{z}|w\rangle\langle\bar{w}|u\rangle 
\end{equation*}
where  $\alpha = -\frac{i\pi}{4}$. Also note that we have inserted the resolution of identity (\ref{A68}) here. Now using 
$$\langle\bar{z}|[\hat{b},\hat{a}]|w\rangle = \partial_{\bar{z}}a(\bar{z},w)-w a(\bar{z},w)$$
$$\langle\bar{z}|[\hat{b}^{\dagger},\hat{a}]|w\rangle = \bar{z}a(\bar{w},z)-\partial_w a(\bar{z},w)$$
and writing $\langle\bar{w}|\hat{a}|z\rangle = a(\bar{w},z) = f(\bar{w},z)e^{\bar{w}z}$ (for  normal ordered operator $\hat{a}$, see (\ref{poly}) ), we can further simplify the above equation as,

\begin{equation}
\int \, d\mu(\bar{w},w) e^{\bar{z}w+\bar{w}z} \left(e^{\alpha}\partial_{\bar{z}}f(\bar{z},w)-e^{-\alpha} \partial_wf(\bar{z},w)\right)\left(e^{-\alpha}\partial_{\bar{w}}f(\bar{w},u)-e^{\alpha}\partial_uf(\bar{w},u)\right) = \theta \int \, d\mu(w,\bar{w}) e^{\bar{z}w+\bar{w}z}
\end{equation}
By comparing the coefficients of $e^{\bar{z}w+\bar{w}z}$ in the integrands of either sides of this equation, we can easily see that  the following set of  differential equations are necessarily satisfied by the function $f(\bar{z},w)$:
\begin{align}
e^{\alpha} \partial_{\bar{z}}f(\bar{z},w)- e^{-\alpha}\partial_w f(\bar{z},w)& = \sqrt{\theta}\,\, pe^{\lambda}\nonumber\\
e^{-\alpha} \partial_{\bar{w}}f(\bar{w},u)-e^{\alpha}\partial_uf(\bar{w},u) &= \sqrt{\theta}\,\, p^*e^{-\lambda}\label{D1}
\end{align}
where $p\in\,\mathbb{C}$ with $|p|=1$  and $\lambda\in\,\mathbb{R}$. This ensures that the product of these two expressions, occurring in the left hand side of this pair of equations is indeed $\theta$. We now replace the variable $\bar{w}$ by $\bar{z}$ and $u$ by $w$ in the second equation of (\ref{D1})  and write the equations in matrix form as,
\begin{equation}
\begin{pmatrix}
e^{\alpha}&-e^{-\alpha} \\
e^{-\alpha}&-e^{\alpha}
\end{pmatrix} \begin{pmatrix}
\partial_{\bar{z}}f\\ \partial_w f
\end{pmatrix} = \sqrt{\theta} \begin{pmatrix}
pe^{\lambda}\\ p^*e^{-\lambda}
\end{pmatrix}
\end{equation}
Inverting the above coefficient matrix and solving for $\partial_{\bar{z}}f$ and $\partial_wf$ we get,
\begin{align}
\partial_{\bar{z}}f &= \frac{\sqrt{\theta}}{2} e^{\frac{i\pi}{4}}\left(pe^{\lambda}-ip^* e^{-\lambda}\right) =K_1\nonumber\\
\partial_w f&= \frac{\sqrt{\theta}}{2} e^{\frac{i\pi}{4}}\left(ipe^{\lambda}-p^*e^{-\lambda}\right) =K_2\label{E2}
\end{align}

So  from the above equations we can infer that  $f(\bar{z},w)=K_1\bar{z}+K_2 w$, which in turn gives the form of  $\hat{a}$  as, $\hat{a}= K_2 \hat{b}+K_1\hat{b}^{\dagger}$. Now imposing the hermiticity condition on the algebra element $\hat{a}$, we simply arrive on the conclusion that $K_1=\bar{K}_2$.Equating the real and imaginary part of this equation we find  $p=i$.  So finally making use of this in (\ref{E2}) , we get a one parameter family of $\hat{a}$ dependent on $\lambda\,\in\,\mathbb{R}$ to be  given by,
\begin{equation}
\hat{a}_{\lambda}=\sqrt{\frac{\theta}{2}} [\hat{b}(\cosh\lambda-i\sinh\lambda)+\hat{b}^{\dagger}(\cosh\lambda+i\sinh\lambda)] = \hat{t}\cosh\lambda+\hat{x}\sinh\lambda
\end{equation}
which is (\ref{aform2}) itself.
We now emulate the commutative case and   substitute this algebra element (\ref{aform2}) in (\ref{ball9}) to get,
\begin{equation}
\textbf{B}[\psi_1,\psi_1^*,\psi_2,\psi_2^*] = -e^{-\lambda}\int d\mu(z,\bar{z}) \psi_1^*(z)\psi_1(\bar{z})-e^{\lambda}\int d\mu(z,\bar{z})\psi_2^*(z)\psi_2(\bar{z})+i\int d\mu(z,\bar{z})[\psi^*_2(z)\psi_1(\bar{z})-\psi_1^*(z)\psi_2(\bar{z})]\label{kol}
\end{equation}
Using  scaling transformation  $\psi_1(\bar{z})\to \psi_1^{\prime}(\bar{z}):= e^{-\frac{\lambda}{2}}\psi_1(\bar{z});\,\, \psi_2(\bar{z})\to \psi_2^{\prime}(\bar{z}):=e^{\frac{\lambda}{2}}\psi_2(\bar{z})$, as before we rewrite (\ref{kol}) as,
\begin{equation}
\textbf{B}= -\int d\mu(z,\bar{z})[\psi_1^{\prime *}(z)\psi_1^{\prime}(\bar{z})+\psi_2^{\prime *}(z)\psi_2^{\prime}(\bar{z})+i\{\psi_1^{\prime *}(z)\psi_2^{\prime}(z)-\psi_2^{\prime *}(z)\psi_1^{\prime}(z)\}]
\end{equation}
It seems that the functions on which \textbf{B} depends are not all independent. So we again  use a transformation similar to (\ref{trans5}) as
\begin{equation}
\begin{pmatrix}
\psi_1(\bar{z})\\ \psi_2(\bar{z})
\end{pmatrix}\to
\begin{pmatrix}
\phi_+(\bar{z})\\ \phi_-(\bar{z})
\end{pmatrix}
=\frac{1}{2}\begin{pmatrix}
1&i \\ -1&i
\end{pmatrix}
\begin{pmatrix}
\psi_1^{\prime}(\bar{z})\\ \psi_2^{\prime}(\bar{z})
\end{pmatrix}
\end{equation}
to get  \textbf{B}  as $\textbf{B}[\phi_-,\phi_-^*]$ in the following way,
\begin{equation}
\textbf{B}= -2\int d\mu(z,\bar{z}) |\phi_-(z)|^2
\end{equation}
Above equation clearly shows, \textbf{B} indeed is independent of $\phi_+$ and $\phi_+^*$ and  \textbf{B} reaches its maximum value for $|\phi_-| =0$ which is its supremum value with above choice of $\hat{a}$ (\ref{aform2}). The fact that this is indeed the right choice is re-inforced by the fact that analogous to the Euclidean case discussed in section 2, here too we can find a suitable bound  to the distance function, except that it has to be lower bound in this case. We shall obtain this bound now in the following sub-section and show that this bound is indeed saturated  by the choice of $\hat{a}$ (\ref{aform2}).
\subsubsection{Calculation of the lower bound of distance}
Like in the Euclidean case, here too we can  try to get a bound in the distance between the pure states. Unlike, however,  the previous case here we have to get a lower bound as the distance formula includes computation of infimum. Emulating the same procedure  as (\ref{B1}), we can write,
\begin{equation}
d(\rho_0,\rho_z) = \int_0^1 d\mu \frac{d}{d\mu}[tr(\rho_{\mu z}\hat{a})]= \int_0^1 d\mu \, tr\left(\rho_{\mu z}[\bar{z}\hat{b}-z\hat{b}^{\dagger},\hat{a}]\right)=\frac{1}{\sqrt{2\theta}}\int_0^1 d\mu\,\,tr\left(\rho_{\mu z}\{[\hat{t},\hat{a}](\bar{z}-z)+i[\hat{x},\hat{a}](z+\bar{z})\}\right) \label{C1}
\end{equation}  
We can now parametrize $t$ and $x$ as
\begin{equation}
t = r \cosh\psi;\,\,\, x = r \sinh\psi,\,\,\,\,\,\,\,\textrm{so that}\,\,\,z = \frac{r(\cosh\psi+i\sinh\psi)}{\sqrt{2\theta}}  \label{param}
\end{equation}  
representing  the points on a hyperbola $t^2-x^2=r^2$ in the
forward light cone and  $r$ is a real constant:$0<r<\infty$. Note that with this choice of parametrization we are
restricting the points in the causal cone. After some simplification, (\ref{C1}) can now be re-written as,\begin{equation}
d(\rho_0,\rho_z) = \frac{ir}{\theta}\int_0^1 d\mu \, tr\left(\rho_{\mu z}[\hat{x}\cosh\psi-\hat{t}\sinh\psi,\hat{a}]\right)\label{C4}
\end{equation}
%where in the intermediate step we have considered, without loss of generality, $z$ to be real, as by Lorentz transformation we can always go to a frame where the separation of  two time-like separated  events become purely temporal. In other words, these pair of events can be taken to occur at the same spatial location in a chosen frame. 
But note that, unlike in the Euclidean case (\ref{B2}) we \textit{cannot} just write $tr(\rho_{\mu z}[\hat{x},\hat{a}]) \leq \Vert [\hat{x},\hat{a}]\Vert_{op}$, as here the so-called ball condition has a completely different structure. The ball condition (\ref{ball}), just implies that   the matrix $M:=\mathcal{J}\{[\mathcal{D},\pi(\hat{a})]+i\chi\}$ is a negative operator. Realising its tensor product structure, we can re-write this, by making use of (\ref{adjoint2}) as,  
 \begin{align}
M=\mathcal{J}\{[\mathcal{D},\pi(\hat{a})]+i\chi\} &= -i\begin{pmatrix}
\frac{1}{\theta}[\hat{t}+\hat{x},\hat{a}]& 1\\ 
-1&\frac{1}{\theta}[\hat{x}-\hat{t},\hat{a}]
\end{pmatrix}\nonumber\\
&=-\frac{i}{\theta}[\hat{x},\hat{a}]\otimes\textbf{1}_2+\textbf{1}\otimes\sigma_2-\frac{i}{\theta}[\hat{t},\hat{a}]\otimes\sigma_3 \label{C2}
\end{align}
where $\textbf{1}_2$ is $2\times 2$ identity matrix. Now carrying out a suitable unitary  transformation with $U=U_1\otimes \textbf{1}_2$, we can  diagonalise the hermitian operator  $-\frac{i}{\theta}[\hat{t},\hat{a}]$, such that,
\begin{equation}
U_1^{-1} (-\frac{i}{\theta}[\hat{t},\hat{a}])U_1= \textrm{diag}(\kappa_1,\kappa_2,\cdots ):= \Gamma\,\,;\,\,\,\,\, \kappa_i \in\mathbb{R}\,\,\,\forall \,\,i
\end{equation}
so that $\Gamma$ is possibly an infinite dimensional diagonal matrix and 
\begin{equation}
M\to M^{\prime}= U^{-1}MU=U^{-1}_1(-\frac{i}{\theta}[\hat{x},\hat{a}])U_1\otimes \textbf{1}_2+
\textbf{1}\otimes \sigma_2+\Gamma\otimes\sigma_3\label{E5}
\end{equation}
Now for each entry in the left slot of this total matrix $M^{\prime}$, say the (i,j)-th element, there is a u(2) Lie algebra element and can be written as,
\begin{equation}
\left[U^{-1}_1(-\frac{i}{\theta}[\hat{x},\hat{a}])U_1\right]_{ij}\textbf{1}_2+\delta_{ij}\sigma_2+\Gamma_{ij}\sigma_3\in u(2)\label{E4}
\end{equation} 
This can now be subjected to another SU(2) transformation $U=\textbf{1}\otimes U_2$ with $U_2= e^{\frac{i}{2}\alpha_1\sigma_1}$, with a suitable choice of $\alpha_1$, to bring it to the diagonal form
\begin{equation}
\left[U^{-1}_1(-\frac{i}{\theta}[\hat{x},\hat{a}])U_1\right]_{ij}\textbf{1}_2+\sqrt{(\delta_{ij})^2+(\Gamma_{ij})^2}\,\,\,\sigma_3\,;\,\,\,\,\,\,\Gamma_{ij}=\kappa_i\delta_{ij}\,\,\,\,(\textrm{no sum})\label{E6}
\end{equation}
So that after two successive transformation by $(U_1\otimes \textbf{1}_2)$ and ($\textbf{1}\otimes U_2$), the matrix M (\ref{C2}), as a whole  finally transforms to 
\begin{equation}
M\to M^{\prime\prime}:= (U_1\otimes U_2)^{-1}M(U_1\otimes U_2) =U_1^{-1} (-\frac{i}{\theta}[\hat{x},\hat{a}])U_1\otimes \textbf{1}_2 +\sqrt{1+\Gamma^{\dagger}\Gamma}\otimes \sigma_3\label{E7}
\end{equation}
Negativity of this operator implies,
\begin{equation}
i\,U_1^{-1} [\hat{x},\hat{a}]U_1\otimes\textbf{1} -\sqrt{\theta^2+U_1^{-1}[\hat{t},\hat{a}]^{\dagger}[\hat{t},\hat{a}]U_1}\,\otimes\,\sigma_3\,\,>0\label{F1}
\end{equation}
Now since   $U_1^{-1}[\hat{t},\hat{a}]^{\dagger}[\hat{t},\hat{a}]U_1$ is a positive operator, we readily obtain, 
\begin{equation}
i  [\hat{x},\hat{a}]\otimes \textbf{1} > \theta \otimes\,\sigma_3
\end{equation} 
which by comparing two sides, further implies that
 \begin{equation}
i  [\hat{x},\hat{a}]\, >\, \theta\label{C3}
\end{equation}
As an off-shoot, we observe that the saturation condition of the inequality in  (\ref{F1}), (in case we include the equality sign also in (\ref{F1})) will be satisfied by only those algebra elements which commute  with $\hat{t}$. This implies, in turn, that $\hat{a}(\hat{t},\hat{x})$ depends on $\hat{t}$ only, so that $[\hat{t},\hat{a}(\hat{t},\hat{x})]=0$. The inequality  (\ref{C3}) then implies that $\hat{a}(\hat{t},\hat{x})$ should then be  linear in $\hat{t}$, so that $i[\hat{x},\hat{a}]$ is just $\theta \textbf{1}$. Thus (\ref{C3}) fixes the scale also. One can notice that this feature was absent in the Euclidean case.\\ \\
At this stage, we observe that this inequality along with the above mentioned observations should be respected in all Lorentz frames, as the original  ball condition (\ref{pbp}) is a Lorentz invariant quantity. Using the Lorentz boost transformation (\ref{E1}), we can therefore write    
\begin{equation}
i[\hat{x}^{\prime},\hat{a}]= i[\hat{t}\sinh\phi+\hat{x}\cosh\phi,\hat{a}] \,\,>\,\,\theta \,\,\,\,(\forall\,\,\phi)
\end{equation}
Now by choosing $\phi = -\psi$ and putting the condition in (\ref{C4}) we have the following lower bound for the Lorentzian distance:
\begin{equation}
d(\rho_0,\rho_z) > r =\sqrt{\theta}\sqrt{z^2+\bar{z}^2}\label{E3}
\end{equation}

Now  for evaluation of distance just like Euclidean Moyal plane, here too the pure states are represented by normalized density matrices . The  distance between a pair of pure states (represented in terms of  normalised Glauber-Sudarshan coherent state basis (\ref{coh}) as) $\rho_0=|0\rangle\langle 0|$ and $\rho_z=|z\rangle\langle z|$ representing two time-like separated events $(0,0), (t_1,x_1)$ in forward light cone with $\rho_0\prec\rho_z$, will be given by,
\begin{align}
d(\rho_0,\rho_z)&=\inf_{\hat{a}\in\mathcal{B}}\{[\rho_z(\hat{a})-\rho_0(\hat{a})]^+\} =\inf_{\lambda} \{[\langle z|\hat{a}_{\lambda}|z\rangle -\langle 0|\hat{a}_{\lambda}|0\rangle ]^+\} \nonumber\\
&=\inf_{\lambda}  \sqrt{\frac{\theta}{2}}\{[(z+\bar{z})\cosh\lambda-i(z-\bar{z})\sinh\lambda]^+\}\label{G1}
\end{align}
where again we have defined $z=\frac{t_1+ix_1}{\sqrt{2\theta}}$ in  the spirit of  (\ref{basis}) in section 2 and made use of the form of $\hat{a}_{\lambda}$ given in (\ref{aform2}). Now the infimum of the set is attained by  minimising the set with respect to $\lambda$ which immediately yields the desired distance between two pure states as,
\begin{equation}
d(\rho_z,\rho_w)=\sqrt{\theta}\sqrt{z^2+\bar{z}^2}= \sqrt{t_1^2-x_1^2},\label{dist2}
\end{equation}
showing that  (\ref{dist2}) indeed agrees with the lower bound (\ref{E3}) .\\ \\
Finally, note that the minimisation value of $\lambda$, for which the minimum value of the R.H.S  of (\ref{G1}) is attained  here is given by $\lambda=\tilde{\lambda}$, which satisfies  $\tanh\tilde{\lambda} = \frac{i(z-\bar{z})}{z+\bar{z}}$. Now putting this value of $\lambda=\tilde{\lambda}$ in the expression of $\hat{a}_s$ (\ref{aform2}) we have,
\begin{equation*}
\hat{a}_s = r\cosh\psi [\hat{b}(\cosh\psi+i\sinh\psi)+\hat{b}^{\dagger}(\cosh\psi-i\sinh\psi)]
\end{equation*} 
where we have used the same parametrisation for $z$, as in (\ref{param}). Now choosing $\psi=-\phi$ again, we have, using (\ref{E1},\ref{LT}),
\begin{equation}
\hat{a}_s= r\cosh\phi \,\,(\hat{b}_{\phi}+\hat{b}^{\dagger}_{\phi}) = \sqrt{\frac{2}{\theta}}(r\cosh\phi)\, \hat{t}^{\prime}
\end{equation}
With this form of $\hat{a}_s$ we can explicitly show that  in a Lorentz transformed frame (\ref{E1}), we have $[\hat{t}^{\prime},\hat{a}_s] = 0$ and $i[\hat{x}^{\prime},\hat{a}_s]=\theta \textbf{1}$, which in turn shows that saturation of the condition (\ref{F1},\ref{C3}) is truly reached at the same time when infimum of the distance functional is reached in a specific Lorentz frame, with our choice of $\hat{a}_s$ (\ref{aform2}). We can therefore really identify the R.H.S of  (\ref{dist2}) as the true Lorentzian distance between the pure states $\rho_0$ and $\rho_z$ with $\rho_0 \prec \rho_z$. \\ \\
 As the distance is ISO(1,1) invariant (\ref{ISO}), we can make a suitable translation to give  distance between $\rho_z$ and $\rho_w$  where $\rho_z \prec \rho_w$ as,
\begin{equation}
d(\rho_z,\rho_w) = \sqrt{\theta}\sqrt{ (w-z) ^2+(\bar{w}-\bar{z})^2}= \sqrt{(t_2-t_1)^2-(x_2-x_1)^2}\label{E7}
\end{equation}
where two pure states $\rho_z$ and $\rho_w$  represents the points  $(t_1,x_1) ,(t_2,x_2)$ in forward light cone. Like the Euclidean case, (\ref{E7}) also does not depend on the non-commutative parameter and it  mimics the result of normal geodesic (here straight line) distance which coincides with the flat commutative 1+1 dim space-time.\\ \\
Before we conclude this pen-ultimate section, we would like to make two pertinent observations. Firstly, like in the Euclidean case, here too we are forced to consider extended multiplier algebra where all the family members of (\ref{aform2}) belongs to this .  Secondly, note that we have dealt with the non-unital algebra $\mathcal{A}=\mathcal{H}_q$  for both the Euclidean and Lorentzian Moyal plane. However, as mentioned in the beginning  of section 3 that the fundamental  symmetry $\mathcal{J}$ requires to satisfy $[\mathcal{J},a]=0, \,\forall\,\,  a\in\tilde{\mathcal{A}}$, where $\tilde{\mathcal{A}}$  is the unitized version of $\mathcal{A}=\mathcal{H}_q$. This may give rise to the possibility  that the bounded-ness of the operator $[\mathcal{D},\pi (a)]$ along with the compact-ness of the operator $\pi (a)(\mathcal{D}-\lambda \textbf{1})^{-1}$ for $a=\textbf{1}$,  may get violated  without affecting the metric aspect- as happens in the Euclidean case (see \cite{Mart,MA} for details). One needs to check this aspect with our skew Krein self-adjoint Dirac operator $\mathcal{D}$ as well. We hope to address this issue in a future publication.

\section{Conclusion and Future Plan}
We have shown in this work that the axiomatic formulation provided in \cite{Franco1,Franco4,Eckstein}, serves our purpose adequately and enables us to compute the spectral distance between pair of time-like separated pure states constructed out of Glauber-Sudarshan coherent states on Lorentzian Moyal plane. This computation was made very transparent by making use of un-normalised Fock-Bergman coherent states so that representation of $|.\rangle$  states and its dual $\langle .|$ states are represented by  anti-holomorphic  and holomorphic functions respectively  occurring naturally  in our analysis involving Hilbert-Schmidt operators.\\ 
This exercise suggests that we can have our trust in this construction of spectral triple for 2D Lorentzian Moyal plane, as this reproduces the expected results. To proceed further beyond this point we definitely need to replace our left module by a bi-module, so that fluctuations in the Dirac operator can be constructed.  This should  enable us to introduce prototypes of gauge and Higgs fields. As a first step one may consider doubled Lorentzian Moyal plane for example as a sequel of doubled Euclidean Moyal plane as considered in \cite{Mart,KK}. This should pave the way for realistic model building where almost commutative spaces are upgraded  to fully non-commutative spaces, so as to provide a glimpse towards Planck scale physics.

\section{Acknowledgement}
We would like to thank  Prof. Debashish  Goswami  and Prof. Jyotishman Bhowmick for their useful discussions regarding this paper.
\section{Appendix - A}
\textbf{Fock-Bergman Coherent state basis and their properties:}\\
Here we introduce some of the working formulae and notations  for  the Fock-Bergman  coherent state basis.  Unlike Glauber-Sudarshan  coherent states the Fock-Bergman basis is un-normalised. We define it as,
\begin{equation}
|z\rangle_B = e^{z\hat{b}^{\dagger}}|0\rangle,\,\,\,_{B}\langle\bar{z}| =\langle 0| e^{\bar{z}b},  \textrm{giving}  \,\,\,_{B}\langle\bar{z}|w\rangle_B =e^{\bar{z}w}\label{A67}
\end{equation}
and is related to the normalised Glauber-Sudarshan coherent states (\ref{coh}) as $|z\rangle = e^{-\frac{|z|^2}{2}}|z\rangle_B$. The advantage of using Fock-Bergman basis is that the representation of vectors belonging to $\mathcal{H}_c$ (\ref{k}) (respectively the dual space $\tilde{\mathcal{H}_c}$) are given by  anti-holomorphic $\psi(\bar{z}):=_{B}\langle \bar{z}|\psi\rangle$ (resp. holomorphic $\psi^*(z):=\langle\psi|z\rangle_B$)  functions. To avoid cluttering of indices we shall drop the $B$ subscript for denoting  a Bergman-Fock basis here onwards. The completeness relation is given by 
\begin{equation}
\int d\mu(z,\bar{z})\,\, |z\rangle \langle \bar{z}| =\textbf{1},\,\,\,\textrm{where}\,\, d\mu(z,\bar{z})= e^{-|z|^2}\frac{Re(z)Im(z)}{\pi}\label{A68}
\end{equation}
The overlap of this basis with a Fock state in (\ref{k}) is given by 
\begin{equation}
\langle\bar{z}|n\rangle = \frac{\bar{z}^n}{\sqrt{n!}}; \,\,\, \langle n|z\rangle =\frac{z^n}{\sqrt{n!}}
\end{equation}
Another pair of important identities are,
\begin{equation}
\int d\mu(z,\bar{z}) \langle\psi|z\rangle\langle\bar{z}|w\rangle  =\langle\psi|w\rangle = \psi^*(w),\,\,\, \int d\mu(z,\bar{z}) \langle\bar{w}|z\rangle\langle\bar{z}|\psi\rangle =\langle\bar{w}|\psi\rangle = \psi (\bar{w})\label{A69}
\end{equation}
enabling us to identify  $\langle \bar{z}|w\rangle = e^{\bar{z}w} = \delta^2(\bar{z},w)$  as a delta function in this space. Indeed, one can introduce the  notion of a functional derivative here too. To that end consider the functional $F[\psi(\bar{z})]$, representing a map  from the space of anti-holomorphic function $\{\psi(\bar{z})\}$  to complex numbers: $F: \{\psi(\bar{z})\}\,\,\to\,\,\mathbb{C}$\\
The functional derivative $\frac{\delta F[\psi(\bar{z})]}{\delta\psi(\bar{w})}$ is then defined by the condition 
\begin{equation}
\int \frac{\delta F[\psi(\bar{z})]}{\delta\psi(\bar{w})} \phi(\bar{w})d\mu(w) =\left. \frac{dF[\psi(\bar{z})+\lambda\phi(\bar{z})]}{d\lambda}\right\vert_{\lambda =0}\label{A70}
\end{equation}
where $\phi(\bar{z})$  represents a small perturbaton with a strength $\lambda$. One can then easily evaluate the following functional derivatives:
\begin{equation}
\frac{\delta\psi(\bar{z})}{\delta\psi(\bar{w})}=\delta^2(\bar{z},w)\,\,\,\,\textrm{and}\,\,\,\,\frac{\delta\psi^*(z)}{\delta\psi^*(w)}=\delta^2(\bar{w},z)\label{A71}
\end{equation}
where the second one follows from the first one by taking complex conjugation or alternatively  defining analogous functional derivative by introducing appropriate  functionals on the space of holomorphic functions.  

\section{Appendix - B}
\textbf{On finite dimensional matrix solution of  (12) and some related observations}\\
In \cite{Chaoba} it has been shown that there exist a finite dimensional solution of optimal algebra element $\hat{a}_s$ saturating the ball condition $\Vert [\hat{b},\hat{a}]\Vert_{op} \leq \sqrt{\frac{\theta}{2}}$. For computation of  distance between harmonic oscillator states $\rho_{n+1}$ and $\rho_n$, the  finite dimensional  optimal algebra can be taken as 
\begin{equation}
\hat{a}_s=\frac{\sqrt{\theta}}{2\sqrt{2(n+1)}}\left[ |n+1\rangle\langle n+1|-|n\rangle\langle n|\,\right]\label{form3}
\end{equation}
If we compute the operator $[\hat{b},\hat{a}]^{\dagger}[\hat{b},\hat{a}] $ with this finite dimensional form of $\hat{a}_s$, we shall get,
\begin{equation}
[\hat{b},\hat{a}]^{\dagger}[\hat{b},\hat{a}] = \frac{\theta}{8(n+1)}[4(n+1) |n+1\rangle\langle n+1|+(n+2)|n+2\rangle\langle n+2| +n|n\rangle\langle n|\, ]
\end{equation}  
It is evident from the form of  the equation that the right hand side matrix is a diagonal matrix but not proportional to unit matrix. To compute the operator norm we have to take the highest eigenvalue which is $4(n+1)$, giving $\Vert [\hat{b},\hat{a}]\Vert_{op}=\sqrt{\frac{\theta}{2}}$.
So we have shown that a finite dimensional form of $\hat{a}_s$ can also saturate the ball condition. But the point to note is that the functional differentiation only provides information  about   local extrema. It should finally augmented by the task of getting maximal eigenvalue to get the operator norm. In this discrete case operator norm is not translationally invariant as it depends on $n$ and is used to compute the distance between successive harmonic oscillator  states $\rho_n$ and $\rho_{n+1}$, which is also, therefore non-invariant under translation and is  given by
\begin{equation}
d(\rho_n,\rho_{n+1}) = \sqrt{\frac{\theta}{2(n+1)}}
\end{equation} 
See \cite{Chaoba,Mart3} for details.

%\section{Points to be included}
%1. We have not used the equality in the Ball condition as derived by the reference paper.\\2. We have not unitised the algebra.\\3. Any comment on non- hermiticity of Non hermitian Dirac operator.\\4. Comment on $a$ not being element of $\mathcal{H}_q$.\\5. Abstract\\6. reality of its spectrum is compromised.
\end{document}